\documentclass[twocolumn]{aastex631}
\usepackage{natbib}
\usepackage[utf8]{inputenc}
\usepackage{gensymb}
\usepackage{xcolor}
\usepackage{graphicx}
\usepackage{booktabs}
\usepackage{multirow}
\usepackage{numprint}
\usepackage{float}

\newcommand{\U}[1]{\ensuremath{\mathrm{\ #1}}}
\newcommand{\UU}[2]{\ensuremath{\mathrm{\ #1^{#2}}}}
\npthousandsep{,}

\begin{document}

\makeatletter
\newcommand\footnoteref[1]{\protected@xdef\@thefnmark{\ref{#1}}\@footnotemark}
\makeatother

\title{A VERITAS/Breakthrough Listen Search for Optical Technosignatures}

\author[0000-0002-2028-9230]{A.~Acharyya}\affiliation{Department of Physics and Astronomy, University of Alabama, Tuscaloosa, AL 35487, USA}
\author[0000-0002-9021-6192]{C.~B.~Adams}\affiliation{Physics Department, Columbia University, New York, NY 10027, USA}
\author{A.~Archer}\affiliation{Department of Physics and Astronomy, DePauw University, Greencastle, IN 46135-0037, USA}
\author[0000-0002-3886-3739]{P.~Bangale}\affiliation{Department of Physics and Astronomy and the Bartol Research Institute, University of Delaware, Newark, DE 19716, USA}
\author{P.~Batista}\affiliation{DESY, Platanenallee 6, 15738 Zeuthen, Germany}
\author[0000-0003-2098-170X]{W.~Benbow}\affiliation{Center for Astrophysics $|$ Harvard \& Smithsonian, Cambridge, MA 02138, USA}
\author[0000-0002-6208-5244]{A.~Brill}\affiliation{N.A.S.A./Goddard Space-Flight Center, Code 661, Greenbelt, MD 20771, USA}
\author{M.~Capasso}\affiliation{Department of Physics and Astronomy, Barnard College, Columbia University, NY 10027, USA}
\author[0000-0002-1853-863X]{M.~Errando}\affiliation{Department of Physics, Washington University, St. Louis, MO 63130, USA}
\author[0000-0002-5068-7344]{A.~Falcone}\affiliation{Department of Astronomy and Astrophysics, 525 Davey Lab, Pennsylvania State University, University Park, PA 16802, USA}
\author[0000-0001-6674-4238]{Q.~Feng}\affiliation{Center for Astrophysics $|$ Harvard \& Smithsonian, Cambridge, MA 02138, USA}
\author{J.~P.~Finley}\affiliation{Department of Physics and Astronomy, Purdue University, West Lafayette, IN 47907, USA}
\author[0000-0002-2944-6060]{G.~M.~Foote}\affiliation{Department of Physics and Astronomy and the Bartol Research Institute, University of Delaware, Newark, DE 19716, USA}
\author[0000-0002-1067-8558]{L.~Fortson}\affiliation{School of Physics and Astronomy, University of Minnesota, Minneapolis, MN 55455, USA}
\author[0000-0003-1614-1273]{A.~Furniss}\affiliation{Department of Physics, California State University - East Bay, Hayward, CA 94542, USA}
\author{S.~Griffin}\affiliation{WIPAC and Department of Physics, University of Wisconsin-Madison, Madison, WI 53703, USA}
\author[0000-0002-0109-4737]{W.~Hanlon}\affiliation{Center for Astrophysics $|$ Harvard \& Smithsonian, Cambridge, MA 02138, USA}
\author[0000-0002-8513-5603]{D.~Hanna}\affiliation{Physics Department, McGill University, Montreal, QC H3A 2T8, Canada}
\author[0000-0003-3878-1677]{O.~Hervet}\affiliation{Santa Cruz Institute for Particle Physics and Department of Physics, University of California, Santa Cruz, CA 95064, USA}
\author[0000-0001-6951-2299]{C.~E.~Hinrichs}\affiliation{Center for Astrophysics $|$ Harvard \& Smithsonian, Cambridge, MA 02138, USA}\affiliation{Department of Physics and Astronomy, Dartmouth College, 6127 Wilder Laboratory, Hanover, NH 03755 USA}
\author{J.~Hoang}\affiliation{Santa Cruz Institute for Particle Physics and Department of Physics, University of California, Santa Cruz, CA 95064, USA}
\author[0000-0002-6833-0474]{J.~Holder}\affiliation{Department of Physics and Astronomy and the Bartol Research Institute, University of Delaware, Newark, DE 19716, USA}
\author[0000-0002-1432-7771]{T.~B.~Humensky}\affiliation{Department of Physics, University of Maryland, College Park, MD, USA}\affiliation{NASA GSFC, Greenbelt, MD 20771, USA}
\author[0000-0002-1089-1754]{W.~Jin}\affiliation{Department of Physics and Astronomy, University of Alabama, Tuscaloosa, AL 35487, USA}
\author[0000-0002-3638-0637]{P.~Kaaret}\affiliation{Department of Physics and Astronomy, University of Iowa, Van Allen Hall, Iowa City, IA 52242, USA}
\author{M.~Kertzman}\affiliation{Department of Physics and Astronomy, DePauw University, Greencastle, IN 46135-0037, USA}
\author{M.~Kherlakian}\affiliation{DESY, Platanenallee 6, 15738 Zeuthen, Germany}
\author[0000-0003-4785-0101]{D.~Kieda}\affiliation{Department of Physics and Astronomy, University of Utah, Salt Lake City, UT 84112, USA}
\author[0000-0002-4260-9186]{T.~K.~Kleiner}\affiliation{DESY, Platanenallee 6, 15738 Zeuthen, Germany}
\author[0000-0002-4289-7106]{N.~Korzoun}\affiliation{Department of Physics and Astronomy and the Bartol Research Institute, University of Delaware, Newark, DE 19716, USA}
\author[0000-0002-5167-1221]{S.~Kumar}\affiliation{Department of Physics, University of Maryland, College Park, MD, USA }
\author[0000-0003-4641-4201]{M.~J.~Lang}\affiliation{School of Natural Sciences, University of Galway, University Road, Galway, H91 TK33, Ireland}
\author[0000-0003-3802-1619]{M.~Lundy}\affiliation{Physics Department, McGill University, Montreal, QC H3A 2T8, Canada}
\author[0000-0001-9868-4700]{G.~Maier}\affiliation{DESY, Platanenallee 6, 15738 Zeuthen, Germany}
\author{C.~E.~McGrath}\affiliation{School of Physics, University College Dublin, Belfield, Dublin 4, Ireland}
\author[0000-0001-7106-8502]{M.~J.~Millard}\affiliation{Department of Physics and Astronomy, University of Iowa, Van Allen Hall, Iowa City, IA 52242, USA}
\author{H.~R.~Miller}\affiliation{Santa Cruz Institute for Particle Physics and Department of Physics, University of California, Santa Cruz, CA 95064, USA}
\author{J.~Millis}\affiliation{Department of Physics and Astronomy, Ball State University, Muncie, IN 47306, USA}
\author[0000-0001-5937-446X]{C.~L.~Mooney}\affiliation{Department of Physics and Astronomy and the Bartol Research Institute, University of Delaware, Newark, DE 19716, USA}
\author[0000-0002-1499-2667]{P.~Moriarty}\affiliation{School of Natural Sciences, University of Galway, University Road, Galway, H91 TK33, Ireland}
\author[0000-0002-3223-0754]{R.~Mukherjee}\affiliation{Department of Physics and Astronomy, Barnard College, Columbia University, NY 10027, USA}
\author[0000-0002-9296-2981]{S.~O'Brien}\affiliation{Physics Department, McGill University, Montreal, QC H3A 2T8, Canada}\affiliation{Arthur B. McDonald Canadian Astroparticle Physics Research Institute, 64 Bader Lane, Queen's University, Kingston, ON K7L 3N6, Canada}
\author[0000-0002-4837-5253]{R.~A.~Ong}\affiliation{Department of Physics and Astronomy, University of California, Los Angeles, CA 90095, USA}
\author[0000-0001-7861-1707]{M.~Pohl}\affiliation{Institute of Physics and Astronomy, University of Potsdam, 14476 Potsdam-Golm, Germany}\affiliation{DESY, Platanenallee 6, 15738 Zeuthen, Germany}
\author[0000-0002-0529-1973]{E.~Pueschel}\affiliation{DESY, Platanenallee 6, 15738 Zeuthen, Germany}
\author[0000-0002-4855-2694]{J.~Quinn}\affiliation{School of Physics, University College Dublin, Belfield, Dublin 4, Ireland}
\author[0000-0002-5351-3323]{K.~Ragan}\affiliation{Physics Department, McGill University, Montreal, QC H3A 2T8, Canada}
\author{P.~T.~Reynolds}\affiliation{Department of Physical Sciences, Munster Technological University, Bishopstown, Cork, T12 P928, Ireland}
\author[0000-0002-7523-7366]{D.~Ribeiro}\affiliation{School of Physics and Astronomy, University of Minnesota, Minneapolis, MN 55455, USA}
\author{E.~Roache}\affiliation{Center for Astrophysics $|$ Harvard \& Smithsonian, Cambridge, MA 02138, USA}
\author[0000-0001-6662-5925]{J.~L.~Ryan}\affiliation{Department of Physics and Astronomy, University of California, Los Angeles, CA 90095, USA}
\author[0000-0003-1387-8915]{I.~Sadeh}\affiliation{DESY, Platanenallee 6, 15738 Zeuthen, Germany}
\author[0000-0002-3171-5039]{L.~Saha}\affiliation{Center for Astrophysics $|$ Harvard \& Smithsonian, Cambridge, MA 02138, USA}
\author{M.~Santander}\affiliation{Department of Physics and Astronomy, University of Alabama, Tuscaloosa, AL 35487, USA}
\author{G.~H.~Sembroski}\affiliation{Department of Physics and Astronomy, Purdue University, West Lafayette, IN 47907, USA}
\author[0000-0002-9856-989X]{R.~Shang}\affiliation{Department of Physics and Astronomy, Barnard College, Columbia University, NY 10027, USA}
\author{D.~Tak}\affiliation{DESY, Platanenallee 6, 15738 Zeuthen, Germany}
\author{A.~K.~Talluri}\affiliation{School of Physics and Astronomy, University of Minnesota, Minneapolis, MN 55455, USA}
\author{J.~V.~Tucci}\affiliation{Department of Physics, Indiana University-Purdue University Indianapolis, Indianapolis, IN 46202, USA}
\author{N.~Vazquez}\affiliation{Santa Cruz Institute for Particle Physics and Department of Physics, University of California, Santa Cruz, CA 95064, USA}
\author[0000-0003-2740-9714]{D.~A.~Williams}\affiliation{Santa Cruz Institute for Particle Physics and Department of Physics, University of California, Santa Cruz, CA 95064, USA}
\author[0000-0002-2730-2733]{S.~L.~Wong}\affiliation{Physics Department, McGill University, Montreal, QC H3A 2T8, Canada}
\author{J.~Woo}\affiliation{Columbia Astrophysics Laboratory, Columbia University, New York, NY 10027, USA}
\collaboration{9999}{VERITAS Collaboration}

\author[0000-0003-3197-2294]{D.~DeBoer}
\affiliation{Breakthrough Listen, University of California Berkeley, Berkeley, CA 94720, USA}

\author[0000-0002-0531-1073]{H.~Isaacson}
\affiliation{Breakthrough Listen, University of California Berkeley, Berkeley, CA 94720, USA}
\affiliation{University of Southern Queensland, Toowoomba, QLD 4350, Australia}

\author{I.~de Pater}
\affiliation{Department of Astronomy, University of California Berkeley, Berkeley, CA 94720, USA}

\author[0000-0003-2783-1608]{D.~C.~Price}
\affiliation{International Centre for Radio Astronomy Research, Curtin University, Kent St, Bentley WA 6102, Australia}
\affiliation{Radio Astronomy Laboratory, 501 Campbell Hall, University of California, Berkeley, CA 94720, USA}

\author[0000-0003-2828-7720]{A.~Siemion}
\affiliation{Breakthrough Listen, University of California Berkeley, Berkeley, CA 94720, USA}
\affiliation{SETI Institute, Mountain View, CA 94043, USA}
\affiliation{Department of Physics and Astronomy, University of Manchester, UK}
\affiliation{University of Malta, Institute of Space Sciences and Astronomy, Msida, MSD2080, Malta}
\correspondingauthor{G.~M.~Foote}

\begin{abstract}
 The Breakthrough Listen Initiative is conducting a program using multiple telescopes around the world to search for ``technosignatures'': artificial transmitters of extraterrestrial origin from beyond our solar system. The VERITAS Collaboration joined this program in 2018, and provides the capability to search for one particular technosignature: optical pulses of a few nanoseconds duration  detectable over interstellar distances. We report here on the analysis and results of dedicated VERITAS observations of Breakthrough Listen targets conducted in 2019 and 2020 and of archival VERITAS data collected since 2012. Thirty hours of dedicated observations of 136 targets and 249 archival observations of 140 targets were analyzed and did not reveal any signals consistent with a technosignature. The results are used to place limits on the fraction of stars hosting transmitting civilizations. We also discuss the minimum-pulse sensitivity of our observations and present VERITAS observations of CALIOP: a space-based pulsed laser onboard the CALIPSO satellite. The detection of these pulses with VERITAS, using the analysis techniques developed for our technosignature search, allows a test of our analysis efficiency and serves as an important proof-of-principle.
\end{abstract}

\section{Introduction}
\par \textit{{\rm The} search for extraterrestrial intelligence} (SETI) can be defined as the ``theory and practice of searching for extraterrestrial technology or technosignatures'' \citep{2018arXiv180906857W}. Technosignatures are extraterrestrial signals whose only explanation is that they were produced artificially. Examples of potential technosignatures include interstellar radio-based communications \citep{1959Natur.184..844C}, interstellar laser-based communications \citep{1961Natur.190..205S, 2017AJ....153..251T, 2023arXiv230106971Z}, radio and optical leakage from technological civilizations \citep{1978Sci...199..377S, 2010AsBio..10..121S}, infrared emission from Dyson spheres \citep{1960Sci...131.1667D}, spectral evidence for industrial pollutants in exoplanet atmospheres \citep{Wright2018HandbookOfExoplanets}, and physical artifacts deposited within our solar system \citep{1960Natur.186..670B}. Since the founding of the field in the 1950s, there have been numerous searches for these technosignatures using radio, optical, and infrared telescopes, but the fraction of the total parameter space which has been searched remains extremely low \citep{2018AJ....156..260W}.

\par This paper presents the results of a partnership between the \textit{Very Energetic Radiation Imaging Telescope Array System} (VERITAS) Collaboration and the Breakthrough Listen Initiative in a search for pulsed optical laser-based communications. The Breakthrough Listen Initiative\footnote{\label{footnote:breakthrough}\url{https://breakthroughinitiatives.org/}} is currently the foremost technosignature search campaign \citep{Worden:2017, Isaacson:2017}. It began searching for radio technosignatures in 2016 through a  partnership with the Green Bank Telescope and the Parkes Observatory, subsequently adding MEERKAT in 2018. Similarly in the optical band, a partnership with the Automated Planet Finder in 2016 at the Lick Observatory and the Keck Observatory enabled a spectral search for laser-based communication \citep{2017AJ....153..251T, Isaacson:2019, Lipman:2019}. More recently, Breakthrough has partnered with the exoplanet-hunting \textit{Transiting Exoplanet Survey Satellite} (TESS) to search for anomalous stellar lightcurves, and to search targets of interest from the TESS catalog with radio telescopes \citep{Traas:2021, Franz:2022}. Taken together, these partnerships constitute the most comprehensive search for technosignatures thus far \citep{2019BAAS...51g.223G}.

\par Each search for a specific technosignature has benefits and drawbacks, justifying the approach of performing many such searches concurrently. For example, radio-leakage technosignatures emit continuously in every direction, but the inverse-square law and the expected low radio intensity lead to a requirement for radio telescopes which are still in the planning and construction phases, with the full Square Kilometer Array being a notable example \citep{Siemion:2015}.

\par For pulsed optical laser-based communication, the benefit lies in concentrating all of the emitting power into a small angular diameter over nanosecond timescales. These laser pulses could, in principle, be produced with today's technology, and could be easily distinguished from the emitter's host star, without significant dispersion losses. A $3\U{ns}$, $3.7\U{MJ}$ optical laser pulse, collimated at the source using a $10\U{m}$ reflector and observed from a distance of 1000 light years, would appear approximately $10^4$ times as bright as its host star \citep{2001SPIE.4273..119H, 2004ApJ...613.1270H}. Constructing an interstellar communication system based on this technology is not only theoretically possible, but is currently feasible. While these pulses could be bright when observed from within the beam's solid angle, they would occur only over very short timescales. The optical receiver therefore requires a large-aperture mirror with fast photon detectors and associated instrumentation. These requirements are the same as those for atmospheric Cherenkov telescopes (ACTs), which are used to measure nanosecond-timescale Cherenkov emission from cosmic-ray- and gamma-ray-initiated particle showers in the Earth's atmosphere. These telescopes can therefore be used to search for optical laser pulse technosignatures \citep{2001SPIE.4273..161C, 2001AsBio...1..489E, 2005ICRC....5..387H, 2005neeu.conf..307A}.

\par Nanosecond pulsed SETI searches in the blue/UV region of the electromagnetic spectrum specifically are well-motivated, if the background due to cosmic ray events can be removed. The study of cosmic rays has been ongoing for more than a century and is closely tied to the development of modern physics. Cosmic rays are an important and ubiquitous constituent of the Galaxy --- their energy density is similar to that of starlight, Galactic magnetic fields or the cosmic microwave background radiation. Any developing technological civilization would almost certainly study cosmic rays and, if located on a planet with a transparent atmosphere, would very likely use the atmospheric Cherenkov effect to do so. The key technology required for this --- photomultiplier tubes --- has been widely available to our civilization since the late 1930s. Furthermore, Cherenkov telescopes are (by far) the largest \textit{optical} telescopes in the world. The H.E.S.S. II telescope, currently operating, has a remarkable $28\U{m}$-aperture and the field as a whole has been operating 10-meter-class telescopes since the late 1960s. We argue, therefore, that nanosecond pulsed emission in the blue/UV (at the peak of the spectrum of Cherenkov light) represents a preferred search region for SETI, similar to the famous ``water hole'' in the radio band. 
 Other wavelengths, such as near infra-red, might also be a natural choice since they experience less extinction due to dust. However, we also argue that any advanced civilization attempting to communicate with an emerging technological civilization would be aware that Cherenkov telescopes are one of the earliest methods capable of easily detecting signals over interstellar distances and that these observations will occur naturally as a side-project of fundamental physical investigations. 

\par One of the first such searches was conducted using the \textit{Solar Tower Atmospheric Cherenkov Effect Experiment} (STACEE), which re-purposed a New Mexico solar energy research facility for nighttime operations as a wavefront-sampling ACT  \citep{2005ITNS...52.2977G}. STACEE consisted of a field of 64 steerable heliostats, each with $37\UU{m}{2}$ mirror area. Light received at the heliostats was reflected onto two sets of secondary mirrors at the top of a tower before being focused onto a set of 64 photomultiplier tubes (PMTs). This system had a $10-15 \U{photon}\UU{m}{-2}$ sensitivity between 400 and 500\U{nm}, peaking at 420\U{nm}.  The STACEE Collaboration conducted dedicated observations of 187 targets from the HabCat catalog \citep{2003ApJS..145..181T} for 10 minutes each, between January and May 2007, and did not find any evidence for technosignature signals during their observations  \citep{2009AsBio...9..345H}. 

\par Imaging atmospheric Cherenkov telescopes are also designed to detect atmospheric Cherenkov radiation, but differ from wavefront-samplers such as STACEE in that the telescopes are equipped with photomultiplier tube cameras which allow recording of an image of the Cherenkov light flash. The potential use of such imaging ACTs (IACTs) for SETI for technosignature searches was first discussed in the early 2000s \citep{2003ESASP.539...31T}, and an analysis methodology and a single test observation using the Whipple $10\U{m}$ IACT was performed in 2005 \citep{2005ICRC....5..387H}. The importance of the imaging technique is that it provides efficient discrimination between point-like pulsed optical technosignatures and the enormous background of Cherenkov flashes generated by cosmic ray particle showers in the Earth's atmosphere. 

\par The power of IACTs for optical technosignature searches is dramatically improved when multiple telescopes are combined together in an array. An array of physically separated telescopes provides an additional coincidence requirement for the pulses detected by each telescope, combined with the ability to measure parallax. This approach was first developed using the VERITAS array, in an archival search for pulsed optical technosignatures from KIC~8462852 \citep{2016ApJ...818L..33A}. The analysis allowed efficient identification of laser-like events over the background of cosmic ray images. 

\par The VERITAS Collaboration has since partnered with the Breakthrough Listen Initiative to continue the research started with the study of KIC 8462852. This partnership has led to 30 hours of dedicated observations of Breakthrough Listen targets with VERITAS and an analysis of 110 hours of observations from the VERITAS archive of sky regions containing Breakthrough Listen targets. The analysis and first results of this program are reported here.

\section{VERITAS}
VERITAS is an IACT array, designed to detect Cherenkov radiation from particle showers in the Earth's atmosphere and to identify those initiated by gamma-ray photons over the background of those due to cosmic rays. A description of the VERITAS telescopes can be found in \citet{2006APh....25..391H}, and the methods of ground-based gamma-ray astronomy are summarized in \textit{e.g.} \citet{JHolder2021Chapter6WSPCHandbook}. Here we briefly describe those technical aspects of VERITAS most relevant to the search for optical technosignatures.
\begin{figure*}
  \includegraphics[width=\linewidth]{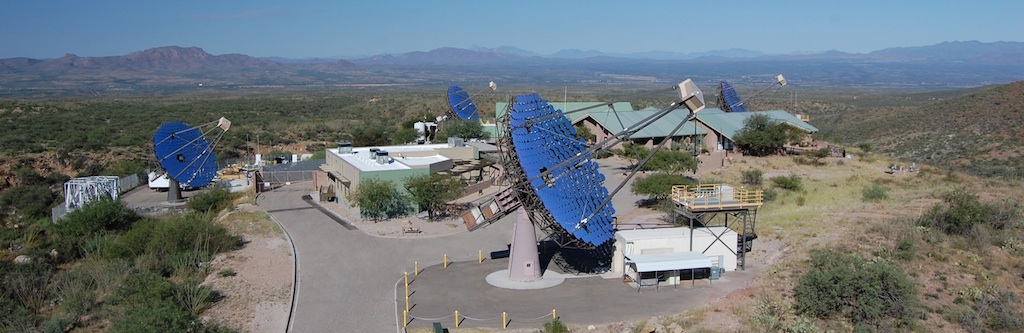}
  \caption{An elevated view of the VERITAS array located at the base of Mount Hopkins near Tuscon, AZ. Pictured are the four individual telescopes, which are roughly $100\U{m}$ apart, the Fred Lawrence Whipple Observatory visitor center, and the VERITAS control building (with the white roof). Image from \citet{2016ApJ...818L..33A}}
  \label{fig:veritas_building}
\end{figure*}
\par VERITAS consists of four IACTs located at the Fred Lawrence Whipple Observatory in southern Arizona  (Figure~\ref{fig:veritas_building}). Each telescope has a 12-m-diameter tessellated reflector mounted on a steerable alt-azimuth platform.  The reflector dish comprises 345 hexagonal mirror segments \citep{2008ICRC....3.1397R} arranged in a Davies-Cotton design \citep{1957SoEn....1...16D}, giving a total mirror area of $\sim110\UU{m}{2}$. Alignment of the individual mirror segments is performed and regularly verified using the method described by \citet{2010APh....32..325M}, resulting in an on-axis optical point-spread function of less than $0.1\degree$ (68\% containment radius).
The focal length of the optical system is $12\U{m}$, giving a focal ratio of 1.0. The focal plane is instrumented with a close-packed array of 499 Hamamatsu R10560 super-bialkali photomultiplier tubes (PMTs), covering an approximately circular  $3.5\degree$-diameter field-of-view (FOV) with a pixel spacing of $0.15\degree$. CCD cameras installed on the telescope structure monitor the position of the PMT camera with respect to the sky, and provide pointing corrections with an absolute positional accuracy of $\sim50\arcsec$. The camera PMT pixels are sensitive over a wide spectral range, with a peak detection efficiency around $400\U{nm}$. Dead space between the circular entrance windows of the PMTs is reduced by the addition of truncated Winston cones to the PMT front faces. These cones are shaped such that the entrance is hexagonal and the exit is circular, allowing them to effectively tile the FOV \citep{2008ICRC....3.1437N}.

All PMT signals are digitized using 2-ns-sampling, 8-bit  flash analog-to-digital converters (FADCs). The FADC read-out is initiated by a three-level trigger system, which requires a signal at the individual pixel level, the telescope camera level, and over the full array. The individual pixel trigger condition is determined by a constant fraction discriminator, while the telescope camera trigger requires at least three adjacent PMT pixel triggers within a coincidence time window of $\sim5\U{ns}$. The array trigger requires at least two telescope camera triggers within a $50\U{ns}$ coincidence window, after the application of hardware timing delays to correct for path-length differences between telescopes. Since the optical technosignature images are expected to resemble the telescope optical point-spread function (which can be smaller than the angular size of a single PMT pixel)  the impact of the 3-pixel camera-level trigger requirement is particularly important for technosignature searches. We discuss this issue in more detail in section~\ref{CALIPSO_section} of this paper.

The recorded FADC pulses are calibrated, integrated and used to create a 499-pixel image for all four telescopes in the array. These images (or ``events") are recorded at a rate of typically $300\U{Hz}$, the vast majority of which are due to Cherenkov emission from cosmic-ray-initiated particle cascades in the atmosphere \citep{2013ICRC...33.1124K}. Subsequent analysis of these images allows identification of the small fraction (typically $<10^{-4}$) that are due to gamma-ray-initiated showers or, as described in the following section, to search for images which resemble a distant optical laser pulse.

\section{OSETI analysis with VERITAS}
\begin{figure*}
    \begin{center}
  \includegraphics[width=0.95\linewidth]{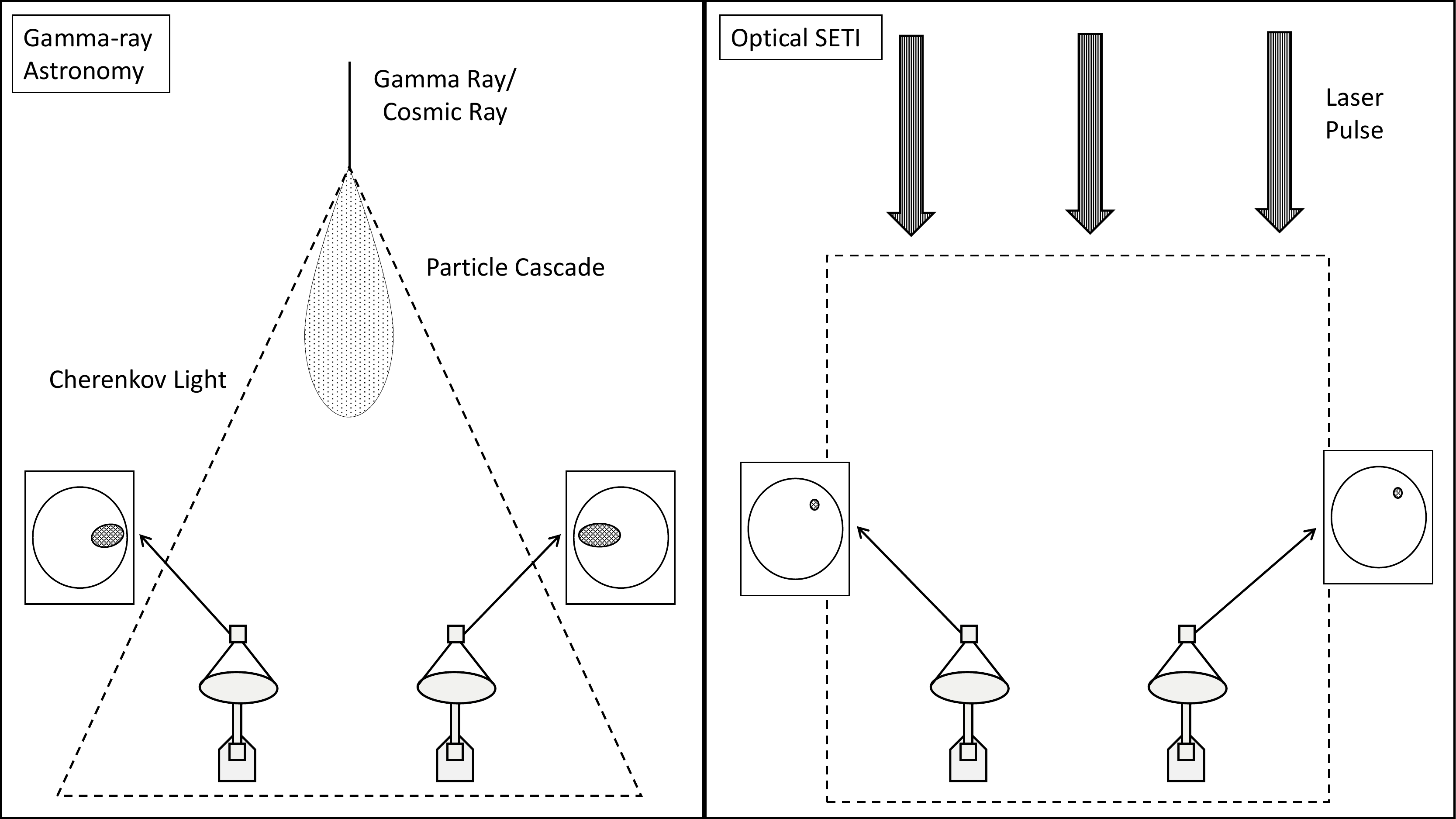}
  \caption{A schematic illustration of the optical SETI (OSETI) technique with IACT arrays such as VERITAS. Particle air showers, initiated by cosmic-ray particles or gamma-ray photons, produce extended images with parallax shifts when viewed from separated telescopes (left). A distant laser pulse produces identical  point-like images, located at the same position in the field of view in each telescope (right).}
  \label{fig:schematic_gamma_seti_astronomy}
  \end{center}
\end{figure*}

\par The analysis applied in this work is similar to that used in the original search for optical technosignatures from KIC\,8462852 with VERITAS \citep{2016ApJ...818L..33A}. The data are first reduced using the standard VERITAS analysis packages \citep{2017ICRC...35..747M, 2008ICRC....3.1385C}, which calibrate and parameterize the recorded images using a moment analysis. Cuts on the resulting image parameters (the image \textit{width}, \textit{length}, etc. --- usually referred to as Hillas parameters \citep{1985ICRC....3..445H}) are then used to filter almost all events due to Cherenkov emission from cosmic ray air showers from the data. While more sophisticated machine learning approaches are under investigation for this analysis, and are already in use for VERITAS gamma-ray analyses \citep{2017APh....89....1K}, simple image parameter cut selections are computationally cheap and robust, and have proven to be extremely effective.

\par The key characteristics of a potential optical technosignature are: (i) that the emission is point-like (i.e. indistinguishable from the telescope optical point-spread function); (ii) that it originates from infinity (i.e. shows no parallax shift and has uniform intensity, when viewed from different locations on the ground); and (iii) that it originates from the position of a target star. This is in contrast to the Cherenkov radiation images of particle cascades produced locally in the Earth's atmosphere, which can have large angular extent (up to a few degrees), are uniformly distributed over the FOV, and which display significant parallax and non-uniform intensity, when viewed by separated telescopes. Although not used in our work, pulse timing differences may also be used to identify technosignature candidates \citep{2018SPIE10702E..5IW}. These features are illustrated schematically in  Figure~\ref{fig:schematic_gamma_seti_astronomy}. 

The choice of image parameter cuts follows logically from these differences. In the analysis of KIC\,8462852, the cuts required that at least three of the four telescope images must contain light, the centroid coordinates of the images in every telescope must be separated from each other by less than $0.15\degree$, and the \textit{length} and \textit{width} of all images must be less than $0.1125\degree$.  After the removal of a few examples of easily identified meteor and satellite tracks, only 28 of the initial 7036970 events passed these selection cuts. There were no events retained in which the average position of the images was within $0.15\degree$ of the location of the target star for that study, KIC\,8462852.
\par In this work, we are analyzing a much larger dataset, testing a catalog of many targets, and have a less homogeneous set of observations. These factors motivate the application of stricter cuts to further improve the background rejection. 
The most important of these is a modification to the \textit{length} and \textit{width} cuts. First, we reduce the cut values to $length<0.09\degree$ and $width<0.07\degree$, this matches the optical point-spread function of the telescopes better than before. Second, we apply these cuts to the telescope with the third-smallest measured \textit{length} or \textit{width} in each event. The motivation behind this is to reduce the impact of PMT afterpulses in the data. Afterpulses are a well-known phenomenon caused mainly by residual positive ions in the PMTs. They appear in the telescope cameras as a single, relatively bright pixel, randomly located in the field of view. This can distort the Hillas parameters of the image in the affected telescope. However, afterpulsing typically affects at most one telescope image in a given event. Applying the \textit{length} and \textit{width} cuts to the image with the third-smallest values of these parameters allows an event with one afterpulse-contaminated image to be retained. The telescope image that exceeds the \textit{length} and \textit{width} cuts is then also removed from consideration for the other selection criteria.
An additional modification is to remove any events that include images potentially truncated by the edge of the camera. This is implemented using the \textit{loss} parameter, defined as the fraction of the total light in the image contained in pixels that lie on the edge of the camera (we require $loss=0$). As a final check, we visually inspect any remaining candidate events (and their associated ancillary data) to ensure that the telescope cameras were functioning correctly, and that each telescope contributed to the event as expected. For example, if a bright pulse was recorded in only three of the four telescopes, this would exclude it as a candidate --- except if the missing telescope had an inoperative PMT pixel at that location in its FOV. At any time, typically a few percent of the PMTs in the telescopes' cameras are malfunctioning, or are temporarily disabled to avoid damage due to bright stars.

\section{Analysis verification using the CALIOP instrument on the CALIPSO satellite}\label{CALIPSO_section}
\par The probability of the VERITAS array triggering on and recording an optical pulse, as well as the efficiency of the subsequent analysis,  is difficult to test under realistic conditions. Monte Carlo simulations provide one approach and are commonly used to estimate the effective detection area and to define the analysis and event selection cuts for gamma-ray astronomy. In this case, the simulated gamma-ray events can be compared with a known bright source of astrophysical gamma-rays, such as the Crab Nebula, and the telescope model parameters tuned until a satisfactory match is achieved. For the \hyphenation{O-SET-I} analysis, however, no natural standard signal exists with which to compare simulations, or to verify the analysis. Furthermore, the precise properties of the pulse to be simulated (risetime, pulse width, wavelength, etc.) are not known. 
\par An ideal test signal would be a distant laser which flashes the telescope cameras from a known location, as this matches the technosignature we are looking for. Pulsed light sources have been used for calibration of IACTs for many years. From 2005, nightly calibrations of VERITAS were performed using a laser with a 337 nm wavelength and a pulse duration of 4 ns at a distance of roughly 5 meters from the camera \citep{2008ICRC....3.1417H} before switching to a similar LED-based calibration system in 2010 \citep{2010NIMPA.612..278H}. However, these measurements are designed to illuminate the entire FOV uniformly and do not serve as a useful analog to a distant point source. Another calibration technique once used by VERITAS involves firing a laser pulse upwards from the ground and observing the Rayleigh-scattered laser light with the telescopes \citep{2005ICRC....5..427S, 2008ICRC....3.1417H}. Again, the observed image is not point-like, but corresponds to an illuminated column in the atmosphere.

The \textit{Cloud-Aerosol Lidar with Orthogonal Polarization} (CALIOP) instrument onboard the polar orbiter \textit{Cloud-Aerosol Lidar and Infrared Pathfinder Satellite Observation} (CALIPSO) satellite is a space-based backscattering lidar, designed to provide high-resolution vertical profiles of aerosols and clouds, which emits $110\U{mJ}$, $20\U{ns}$-duration laser pulses at a repetition rate of $20.16\U{Hz}$, at both $532\U{nm}$ and $1064\U{nm}$ \citep{2009JAtOT..26.2310W}. This provides an excellent technosignature verification source for VERITAS. The camera PMTs are sensitive at $532\U{nm}$, with a quantum efficiency of approximately 12\%. At an orbital height of $700\U{km}$, the laser is effectively a point source at infinity relative to the size of VERITAS. The lidar is directed 3\degree\ from geodetic nadir in the forward along-track direction of the satellite's orbit, and the laser footprint on the ground is predicted to be less than 100 m in diameter \citep{2009JAtOT..26.2310W}, making a coincidental overlap with the VERITAS telescopes extremely unlikely. However, observations by the TAIGA-HiScore collaboration of the CALIPSO laser \citep{2022icrc.confE.876P} demonstrated that the actual footprint extends far beyond the nominal distance, out to at least tens of kilometers, for reasons which are not entirely clear. This motivated both new observations with VERITAS, and a search of the VERITAS archive for serendipitous passages of the satellite through the field of view. 

\begin{figure}[h]
\includegraphics[width=1.0\columnwidth]{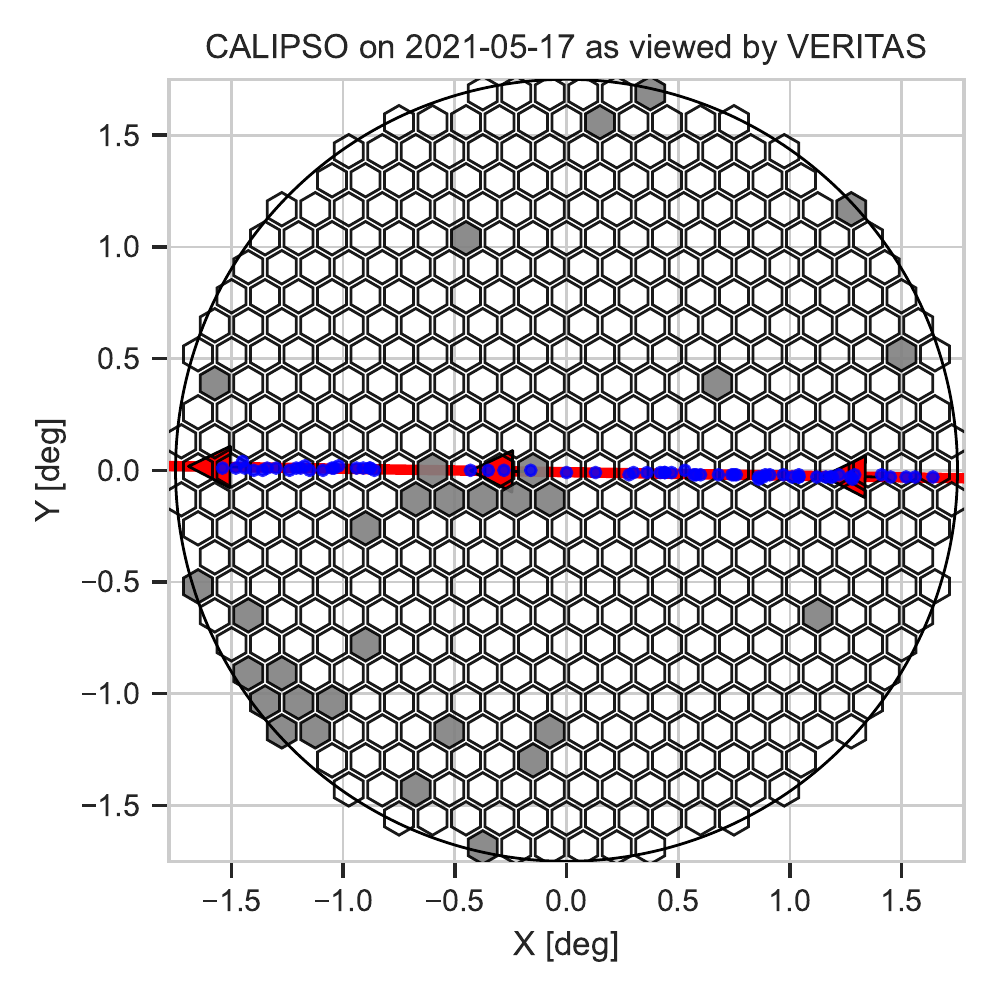}
\includegraphics[width=1.0\columnwidth]{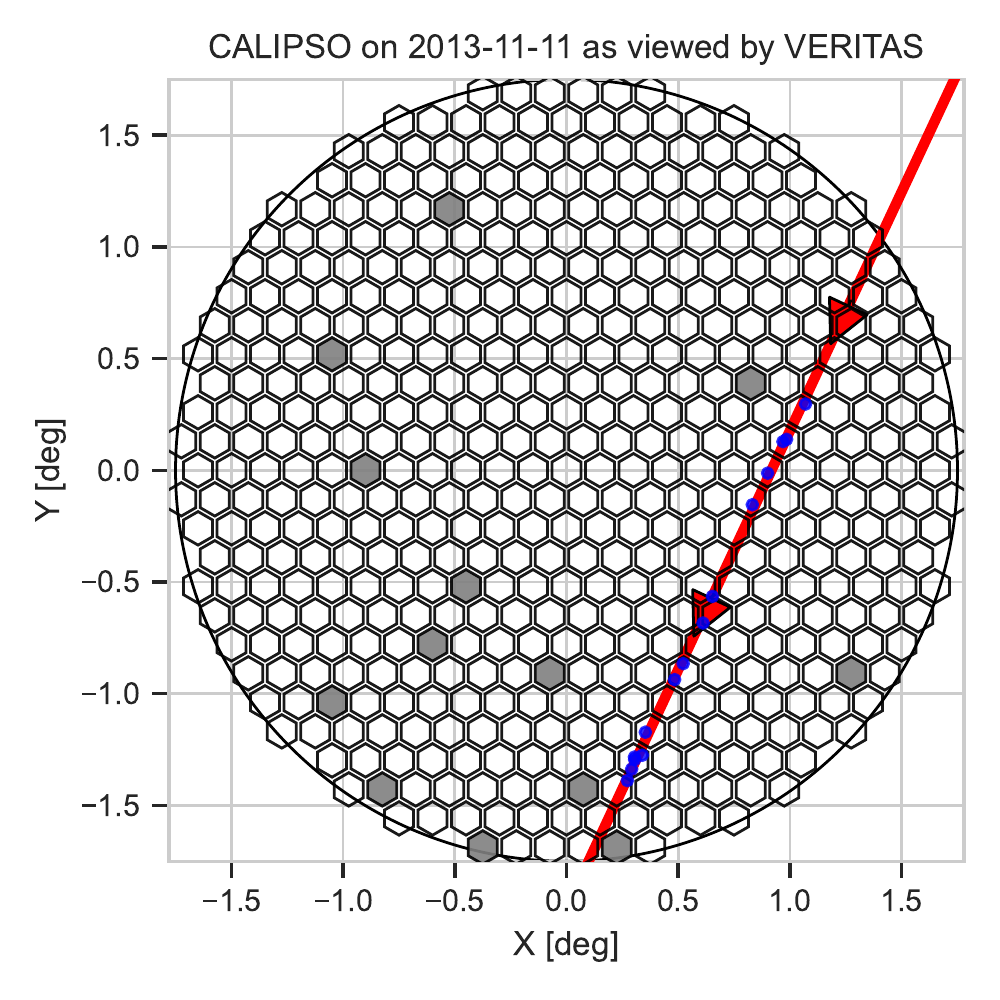}
	\caption{The tracks of the CALIPSO transits (in camera coordinates) on May 17, 2021 \textbf{(top)} and November 11, 2013 \textbf{(bottom)} with the location of the average of the image centers in an event shown as blue points. The hexagons correspond to VERITAS pixels. Grayed out pixels were non-functional during the transit for VERITAS telescope 1.}
	\label{fig:calipso-track}
\end{figure}

Examples of these CALIPSO observations are shown in Figure~\ref{fig:calipso-track}. The top image illustrates a passage from a dedicated observation on May 17, 2021,  which occurred at an elevation of 74\degree. The pulse intensity observed by VERITAS was approximately 2000 photo-electrons at each telescope, corresponding to 150\U{photons}\UU{m}{-2} at ground level. During the transit, 69\% of the pulses emitted by CALIPSO triggered VERITAS and 55\% passed the optical SETI analysis cuts without accounting for the \textit{loss} parameter. The efficiency without the \textit{loss} cut applied is most relevant for comparison with our analysis, since we select target locations which are not close to the camera edge (section \ref{sec:observations}). For these very high intensity pulses, the missing triggered events are largely explained by the deadtime of the telescope data acquisition (which was 9\% for this observation) and by the existence of a large patch of inoperative PMTs in one telescope along the track of the satellite, as can be seen in the figure. 

A second transit is shown on the bottom of Figure~\ref{fig:calipso-track}. This was a serendipitous passage which occurred on November 11, 2013 at an elevation of 54\degree, when the center of the laser footprint was $\sim400\U{km}$ distant from the location of VERITAS. The measured pulse intensity is more than an order of magnitude lower in this case, corresponding to approximately 10\U{photons}\UU{m}{-2}. 21\% of the pulses emitted by CALIPSO triggered VERITAS and 20\% passed our analysis without accounting for \textit{loss}. The relatively low trigger efficiency during this transit is likely the result of the extremely non-uniform sensitivity of the VERITAS trigger system to point-like pulses, as we discuss in more detail in section~\ref{section:discussion}. 

\section{Observations and Results}\label{sec:observations}

The VERITAS/Breakthrough Listen search we have conducted comprises two datasets, which we describe below. The first of these is a program of dedicated VERITAS observations of Breakthrough Listen targets, while the second is an analysis of serendipitous archival observations.

\subsection{Dedicated VERITAS Observations}

Between March 2019 and March 2020, VERITAS spent 30 hours observing objects selected from the Breakthrough Listen  target catalog \citep{2017PASP..129e4501I}. This catalog lists targets originally identified for Breakthrough Listen observations with the Green Bank Telescope, Parkes Telescope, and the Automated Planet Finder. It includes: the 60 nearest stars; 1649 stars within a distance of $50\U{pc}$ sampling a range of masses, ages and elemental abundances; 123 nearby galaxies; and several exotic objects, including white dwarfs, brown dwarfs, neutron stars and black holes.
 
Not all of these targets are suitable for observations with VERITAS. We removed all galaxies, on the assumption that optical emission is unlikely to be detectable over such large (i.e. extragalactic) distances \citep{2018JApA...39...73H}. We also limited targets to the declination band between $\delta=-10\degree$ and $\delta=+70\degree$, to ensure that the object culminates above $\sim40\degree$. High elevation observations are preferred, as they provide greater discrimination power between a pulsed point source at infinity and the background of Cherenkov events generated in the Earth's atmosphere. This is due to the fact that Cherenkov flashes observed at low elevation occur at a larger distance from the telescopes, reducing their parallax angle, image intensity and angular size and making them appear more point-like. 

We also removed all targets with a B-band magnitude of less than 7, and all targets within $0.15\degree$ of an object (excluding the target itself) with a B or V magnitude of less than 8. Bright stars generate a large amount of background photon noise in the VERITAS PMTs, as well as high currents, which accelerate wear. If the current on any individual PMT exceeds a preset threshold, the high voltage supplied to that channel is automatically turned off. This has little impact on the observation of Cherenkov events with large angular extent, but reduces or completely removes the sensitivity to point-like optical pulses from the star's location. 

Finally, we removed any targets which had been previously observed by VERITAS, either intentionally or serendipitously, in the FOV of other observations. Observations of these targets are included in the archival search described in section~\ref{sec:archival}. This resulted in a list of 506 targets, which were then ranked according to the inverse square of their distance and their optical brightness, with nearby, optically faint targets being preferred. Targets lying close to the ecliptic (which could host civilizations that view the Earth as a transiting exoplanet \citep{2016AsBio..16..259H, 2020AJ....160...29S}), or hosting known exoplanets, or located close to another target, were also favored, but with lower weight than the two main criteria of brightness and distance. 

Candidates from this target list were selected for observation based on their ranking and on observatory scheduling constraints. Observations were conducted in typically 15 minute exposures taken within 90 minutes of  culmination, with the primary target offset from the center of the FOV by 1.25\degree. This offset was chosen to improve the probability of triggering on a faint pulse, as discussed in more detail in section~\ref{sec:sensitivity}. All data were taken under clear skies, at new or crescent Moon phases, and with all four telescopes in the array operating correctly. The final dataset comprises 127 observing runs of 108 non-overlapping target fields, with a total exposure of 30.16 hours. Some target fields contain multiple targets, allowing us to study a total of 136 targets with this dataset. Most targets were observed only once, while 25 were observed twice, and three were observed three times.

\begin{figure*}
  \includegraphics[width=\linewidth]{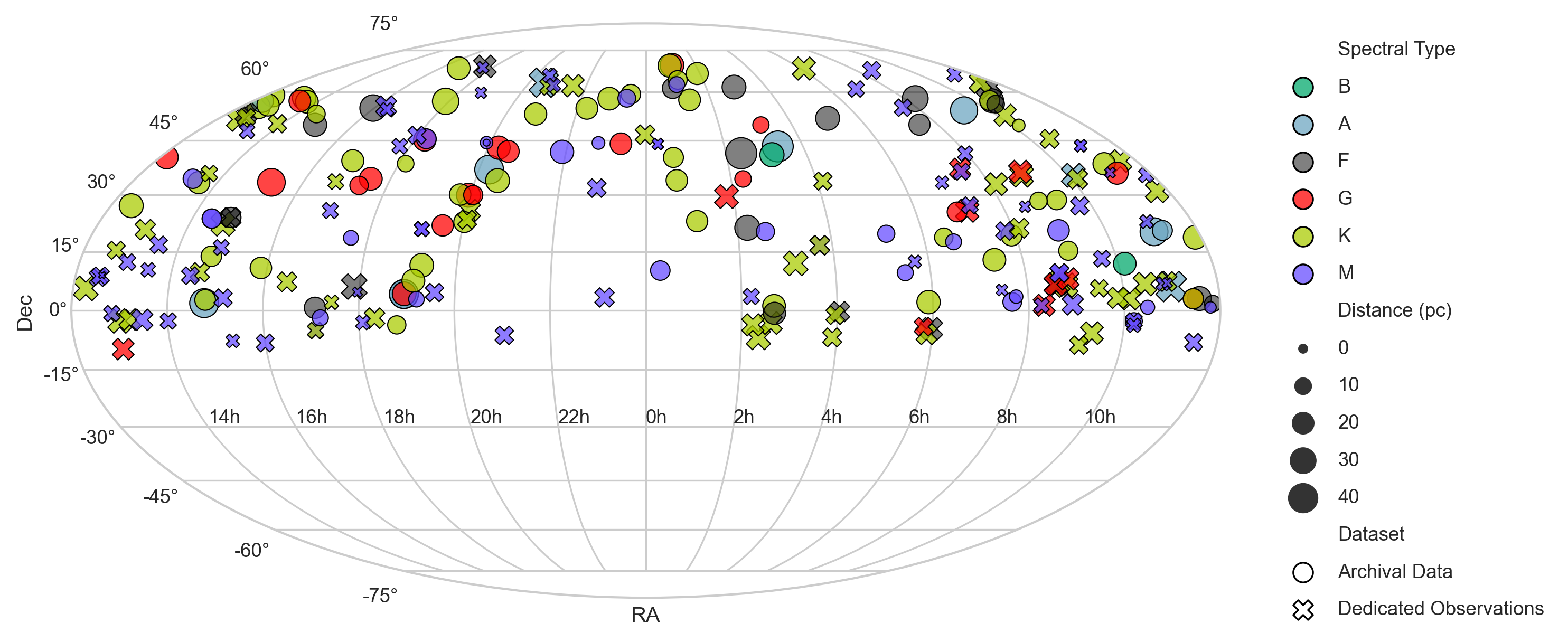}
  \caption{The stellar target locations, in equatorial coordinates, for both the dedicated and archival VERITAS observations. Distance and spectral type are also indicated, as described in the figure legend.}
  \label{fig:skymap}
\end{figure*}

\begin{figure}
  \includegraphics[width=1.0\columnwidth]{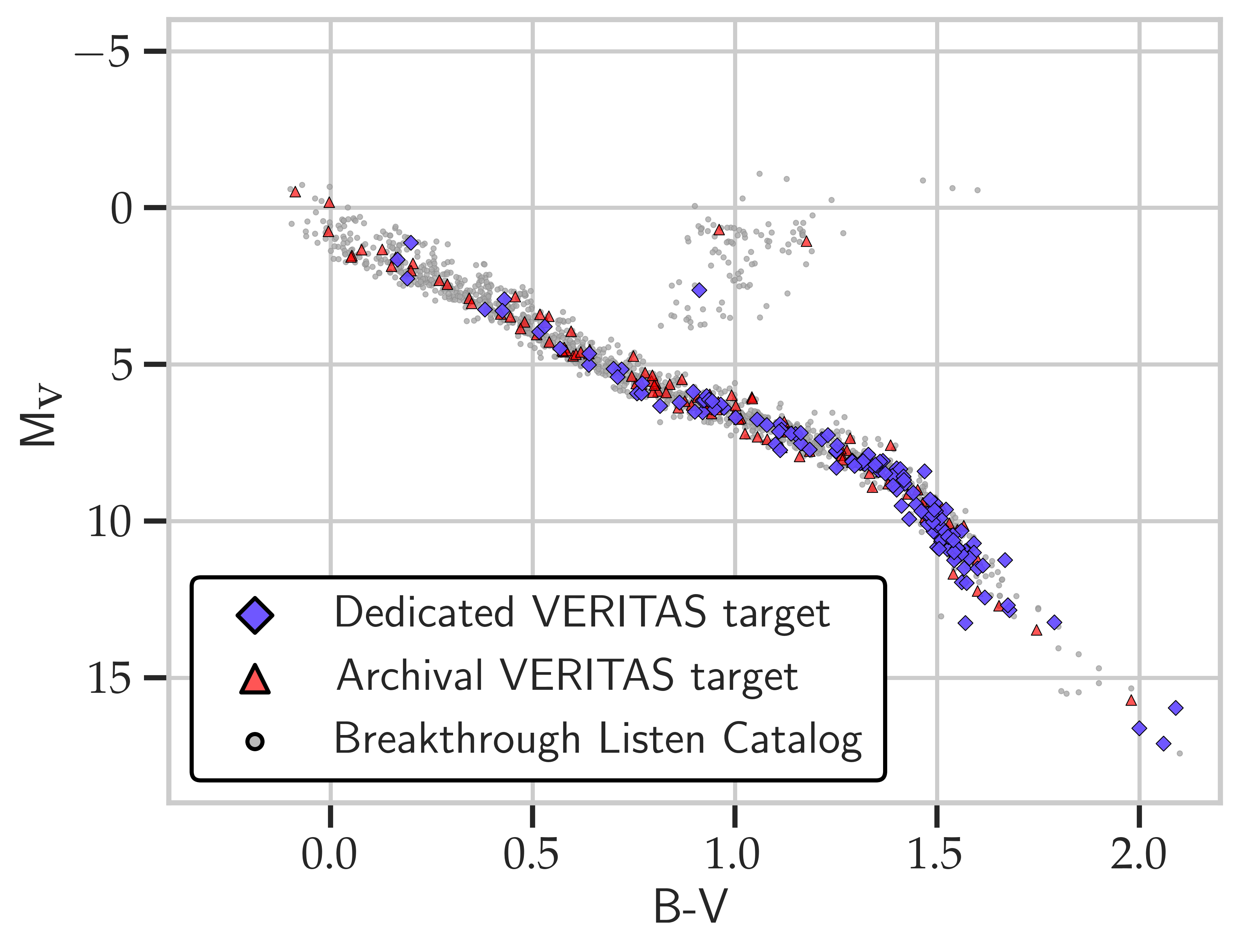}
  \caption{The coverage across an H-R Diagram for the stellar targets used in both dedicated observations and archival analysis. It is similar to the coverage found in the original \citet{2017PASP..129e4501I} Catalog.}
  \label{fig:issacson-hr-fig}
\end{figure}

The locations of the observed stellar targets are shown in figure~\ref{fig:skymap}, with their spectral class and distance indicated. Figure~\ref{fig:issacson-hr-fig} shows the Hertzsprung-Russell diagram for all of the stellar Breakthrough Listen targets, with those targets observed by VERITAS indicated. The VERITAS selection covers a broad range of spectral classes along the main sequence, from B to M, as well as a few giant branch stars.

\subsection{Archival Search}\label{sec:archival}

\par VERITAS has been fully operational since 2007, and records typically $\sim1000\U{hours}$ of observations each year. Thanks to the large VERITAS FOV ($9.6\U{deg^2}$), these observations provide coverage of over 20\% of the sky, with exposures cumulatively ranging from a few minutes to hundreds of hours. A complete search of this extensive archive for optical pulses is a worthwhile subject for future work, but will require further development of the analysis tools -- in particular, to deal with observations recorded at low elevation and to overcome the increased background from examining the entire FOV as opposed to just the region around a set of pre-defined locations.

\par For this work, we selected a reduced set of archival VERITAS observations to analyze. To create this set, we required that the observations were recorded at high elevations ($>40\degree$), with at least three telescopes operating, and with excellent weather conditions. Data taken prior to summer 2012 were not considered, as this was when the VERITAS photomultiplier tube cameras were upgraded, improving the photon detection efficiency by $\sim30\%$ \citep{2013ICRC...33.1124K}. Unlike the dedicated observations, the radial distance of the target from the center of the field of view for archival observations could not be fixed at $1.25\degree$; We instead set the maximum radial distance to be $1.5\degree$. We also set the maximum exposure to be analyzed on any single target to be $1\U{hour}$. If any target exceeded this threshold, we analyzed the first hour of good-quality data and left the remainder for the future full archival analysis. With these conditions, we selected 249 archival observations with an average length of 28 minutes which altogether represents 110 hours of observations containing 140 individual Breakthrough Listen targets and 119 non-overlapping fields. The list of Breakthrough Listen targets in this archival dataset includes 25 galaxies, which were serendipitously inside the studied fields and are included here for completeness. This dataset constitutes all of the Breakthrough Listen targets for which we have good quality, high elevation data taken between September, 2012 and March, 2019. We note that the entire VERITAS archive comprises almost $20,000\U{hours}$ of data, including additional observations of the Breakthrough targets analyzed here. The full analysis of this dataset will be presented in future work. Figure~\ref{fig:skymap} shows the location, spectral type, distance, and originating dataset of all analyzed targets. Figure~\ref{fig:issacson-hr-fig} shows the same targets, but instead within the H-R diagram, showing the color and magnitude of the targets. As the figures show, the analyzed targets selected occupy a significant portion of the parameter space --- locations, distances and spectral properties --- of the \citet{2017PASP..129e4501I} catalog. 

\subsection{Results}
Table \ref{table:obs-cut-res} shows the results of each stage of the analysis pipeline for both the dedicated observations and for the archival dataset. For the dedicated observations, only one event survived the pre-defined selection cuts (target HIP\,83043). For the larger archival dataset, three events survived (targets HIP\,51317, HIP\,93871 and NGC\,4551) and were subjected to visual inspection. For three of these four events, two of the four telescopes in the array triggered and three of the four telescopes registered an image. The fourth telescope did not, despite being operational and having no disabled PMTs at the pulse location. These events therefore fail our requirement for uniform intensity, and are rejected. The remaining event shows three OSETI-like images, thereby satisfying the third-smallest \textit{width} and \textit{length} criterion, but the fourth telescope image contains a bright, extended flash, with an angular (parallactic) displacement from the other images. This clearly identifies the event as being due to a cosmic ray air shower in the atmosphere, and so it is also rejected. We therefore have zero candidate events remaining after the full analysis.

\begin{deluxetable*}{lcc}
\tablewidth{0pt}
\tablenum{1}
\tablecolumns{3}
\tablecaption{The number of events remaining after each stage of the analysis for serendipitous archival observations and for dedicated observations of Breakthrough Listen targets\label{table:obs-cut-res}}
\tablehead{
\colhead{Cut description} & \multicolumn{2}{c}{Events remaining after cut} \\
\colhead{} & \colhead{Archival Data} & \colhead{Dedicated Observations}
}
\startdata
Before cuts & \numprint{127346295} & \numprint{34917340} \\ 
At least 3 images & \numprint{80910174} & \numprint{23088334} \\ 
Point-like images ($3^{\text{\tiny rd}}$ smallest $length<0.09\degree$ \&  $width<0.07\degree$) & \numprint{1894155} & \numprint{508637} \\ 
Image centers co-located (within $0.15\degree$)  & 237 & 35 \\ 
Near target (within $0.15\degree$)  & 3 & 1 \\ 
Images are not truncated ($loss=0$) & 3 & 1 \\
Visual inspection & 0 & 0 \\
\bf{Candidate Events} & \bf{0} & \bf{0} \\
\enddata
\label{table:obs-cut-res}
\end{deluxetable*}

\section{Discussion}\label{section:discussion}
We have presented the analysis of observations of 272 Breakthrough Listen targets with VERITAS (there are 4 targets in common between the target lists of the dedicated observations and the archival search) and have found no evidence for rapid optical pulses from any of these objects. Here we attempt to summarize the sensitivity of our search, both in terms of the minimum optical pulse intensity detectable by VERITAS and in the constraints our survey allows us to place on the frequency of emitting civilizations. 

\subsection{Optical pulse sensitivity}\label{sec:sensitivity}
\citet{2016ApJ...818L..33A} estimated the minimum optical pulse intensity detectable by VERITAS to be $0.94\U{photons}\UU{m}{-2}$ for a $12\U{ns}$ integration window, while noting that such estimates are challenging due to the various unknown pulse properties (location, wavelength, duration, temporal profile, etc). The CALIPSO observations demonstrate experimentally that pulses with an intensity of $10\U{photons}\UU{m}{-2}$ can be detected. Furthermore, the CALIPSO pulses are relatively long duration, with a pulse width of $\sim20\U{ns}$. The single photo-electron pulse width for VERITAS is $4\U{ns}$, and the camera trigger coincidence time is $\sim5\U{ns}$, implying that shorter pulses with a substantially lower integrated photon intensity must also be detectable. However, the CALIPSO results also highlight that the efficiency for pulse detection with VERITAS is not 100\%. We discuss one of the reasons for this in more detail here.

As mentioned, the VERITAS telescope cameras are each composed of 499 photomultiplier tubes on a hexagonal grid, with a pixel to pixel spacing of $0.15\degree$. For an individual telescope to trigger on an optical pulse, signals on three adjacent PMT pixels must exceed a discriminator threshold within a $\sim5\U{ns}$ coincidence window. A laser pulse generated at large distance is point-like, and so this 3-adjacent trigger condition would never be met, if the optical system were perfect. In reality, an image of a point source has the same shape and structure as the telescope optical point-spread function (PSF), which may overlap multiple pixels. The angular extent of Cherenkov showers and the PMT pixels allows for cheaper mirrors with a significantly reduced angular resolution/increased PSF compared to typical optical telescopes. This means that IACTs like VERITAS can be much larger and overall cheaper than their optical counterparts \citep{2010SPIE.7739E..0HC}. The PSF will also vary across the field of view due to comatic aberration. At the center of the camera, it can be approximated by a bivariate Gaussian with a $\sim0.08\degree$ 68\% containment diameter, increasing to $\sim0.15\degree$ at an offset of $1.2\degree$, and degrading further towards the edge of the camera at $1.75 \degree$.

The probability of satisfying the 3-adjacent trigger condition therefore depends very strongly upon the exact pulse location in the field of view. Specifically, it is determined by the amount of light received by the PMT which is third-most-distant from the image centroid --- \textit{i.e.} that which measures the third-largest signal. In the most favorable case, the pulse centroid location lies equidistant between three pixels, each of which receives approximately one third of the light. In the least favorable case, the pulse lands exactly in the center of a pixel, and adjacent pixels receive
only a small fraction of the light. The difference between these two cases is most extreme close to the camera center, where the optical PSF is small, and least extreme at the camera edge, where the optical PSF is more extended.

\begin{figure}
  \includegraphics[width=1.0\columnwidth]{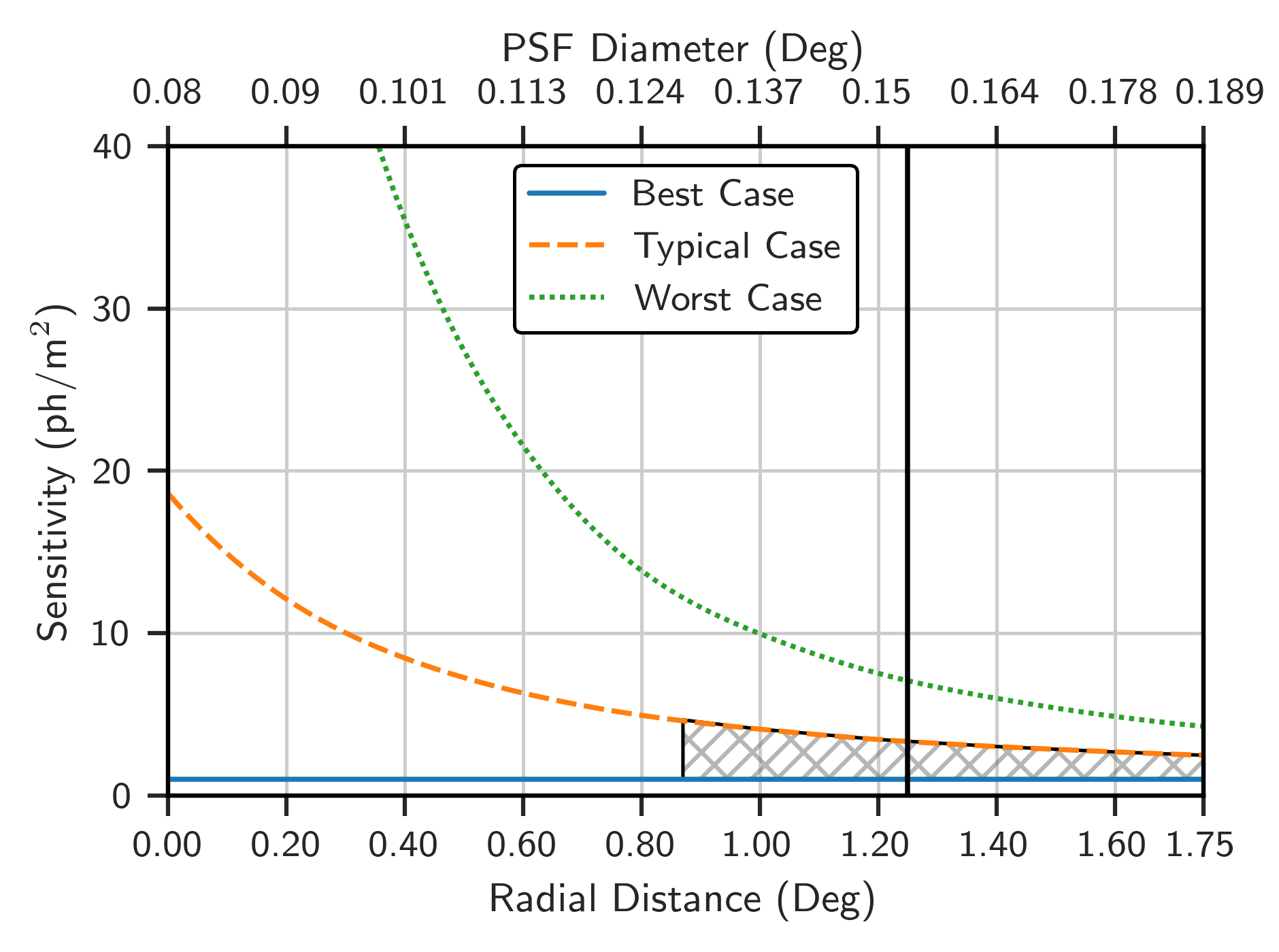}
  \caption{The sensitivity (minimum detectable pulse intensity) as a function of radial distance from the center of the VERITAS telescope field of view. The three curves correspond to a pulse located at the center of a PMT (worst case), equidistant between 3 PMTs (best case) and at the most common location (typical case). The cross-hatched region indicates the outer 75\% of the camera area, and the sensitivity of 75\% of the possible pulse locations within this area. See text for more details.}
  \label{fig:pulse_sensitivity}
\end{figure}

Figure~\ref{fig:pulse_sensitivity} shows the results of a Monte Carlo simulation which illustrates these effects and how the minimum pulse sensitivity varies with radial distance in the camera. The optimum sensitivity is taken to be the same as that estimated by \citet{2016ApJ...818L..33A}. The green dotted line in the figure corresponds to the worst case, where the pulse is centered on a pixel. The blue solid line corresponds to the best case, where the pulse is equidistant between 3 pixels. For a random pulse location on the camera, the most likely distance between the pulse location and the pixel containing the third brightest signal is $0.13\degree$. The orange dashed line indicates this typical case. 75\% of possible pulse locations in the camera  provide a sensitivity equal to or better than this typical case. The cross-hatched region indicates this, as well as the outer region of the camera corresponding to 75\% of the total area. The black vertical line at a camera radius of $1.25\degree$ indicates the position of the Breakthrough Listen targets in the field of view for the dedicated VERITAS observations reported here. The typical sensitivity in this case is $3\U{ph}\UU{m}{-2}$. Observations of the CALIPSO satellite laser over a wide range of elevations (and hence pulse intensity and pulse location in the cameras) are currently being taken by VERITAS and will allow testing these sensitivity estimates more rigorously. 

As a final point, we stress that the issue of non-uniform sensitivity across the field of view is not intrinsic to the technique; rather, it is a result of the VERITAS trigger system design, which is optimized for gamma-ray astronomy. A dedicated trigger for point-like optical pulses, requiring the same single pixel to cross a trigger threshold on multiple separated telescopes, would completely remove this limitation. This could be implemented in a relatively straightforward manner on existing or future facilities and operate in parallel with the existing Cherenkov trigger system. 

\subsection{Survey sensitivity}\label{sec:survey_sensitivity}
The sensitivity to optical pulses is an important instrumental metric. Complementary to this, however, is the sensitivity of the search as a survey: that is, how do our results constrain the parameter space of potential emitters? There are many different ways to estimate this, usually discussed in the context of searches for radio technosignatures (\textit{e.g.} see \citet{2018AJ....156..260W} and references therein). The most applicable prior works for our purposes are those of \citet{2004ApJ...613.1270H}, \citet{2007AcAau..61...78H}, and \citet{curtis_2013} which discuss the search for nanosecond-duration, pulsed optical emission using optical astronomical telescopes equipped with hybrid avalanche photodetectors or with photomultiplier tubes. From 1998 to 2003, 11,600 targeted observations of 4730 stellar objects were made under good conditions with the $1.5\U{m}$-aperture Wyeth telescope at the Harvard/Smithsonian Oak Ridge Observatory, with a total exposure of $1721\U{hr}$. Subsequently, the Harvard All-Sky Observatory utilized a custom optical setup consisting of a $1.8\U{m}$ telescope which focused a $1.6\degree \times 0.2\degree$ patch of the sky onto a beam splitter with matched arrays of 8 photomultiplier tubes down each path. From 2007 to 2012, it made 7320 hours of observations over which it searched the entire northern sky four times. 

\begin{figure}
  \includegraphics[width=\linewidth]{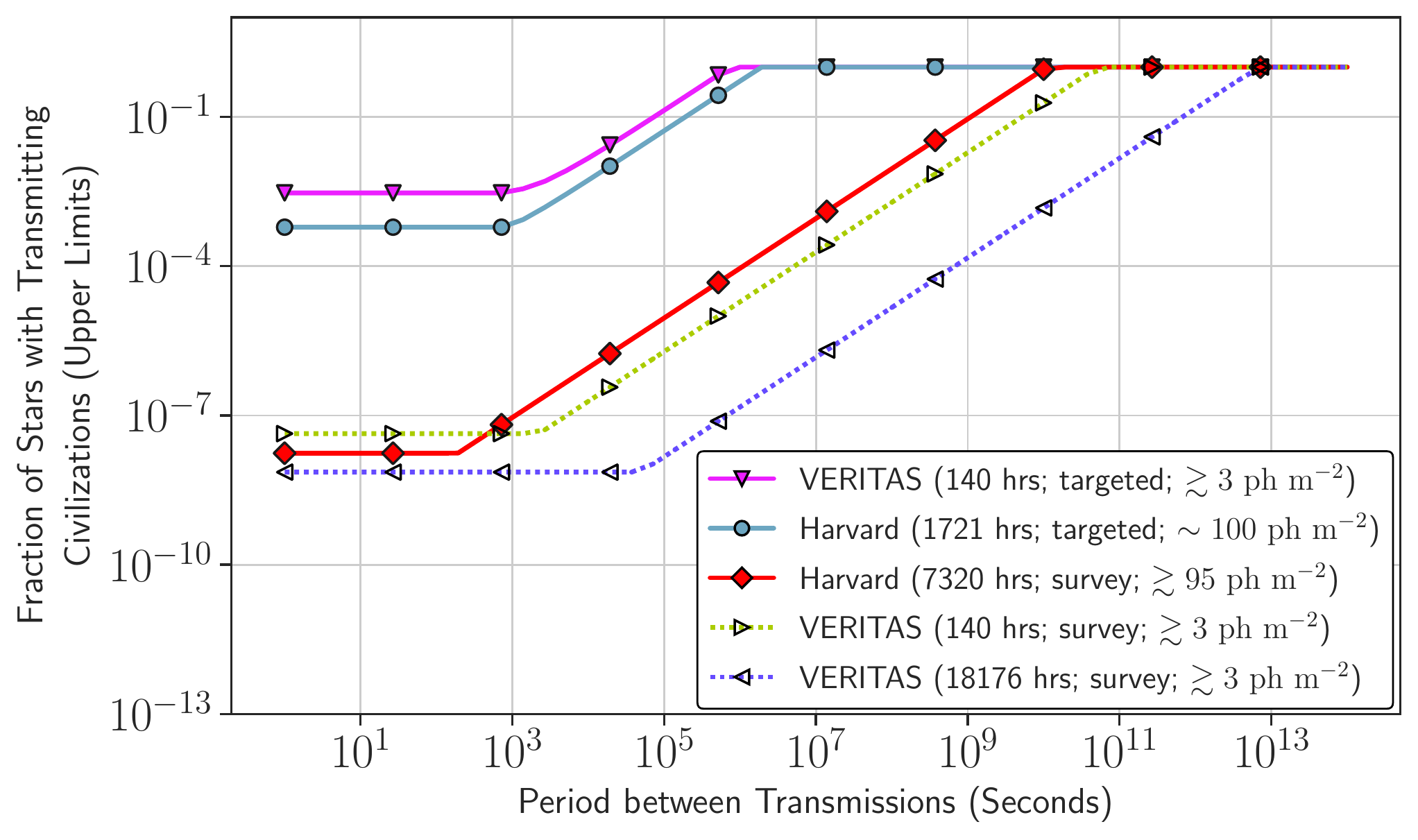}
  \caption{Upper limits on the fraction of stars with transmitting civilizations as a function of the average period between pulses, using the model from \citet{2007AcAau..61...78H} (building on \citet{2004ApJ...613.1270H}) which assumes no candidate pulses are found. From top to bottom the five lines correspond to: all data from this paper (247 targets; upside-down triangles), the Harvard targeted search \citep{2004ApJ...613.1270H}(4730 targets; circles), the Harvard All-Sky untargeted survey \citep{2006PhDT........58H, curtis_2013} (7320 hours; diamonds), a hypothetical non-targeted VERITAS survey using all of the data from this paper (140 hours; right-pointing triangles), and a hypothetical non-targeted survey using all data from the entire VERITAS archive (18,176 hours; left-pointing triangles). The minimum detectable pulse is $\sim 30$ times larger for Harvard than for VERITAS.}
  \label{fig:haystack-fig}
\end{figure}

Each of these campaigns used the same mathematical model, as explained in \citet{2004ApJ...613.1270H}, to place an upper bound on the fraction of nearby stars that host civilizations emitting optical pulses towards the Earth as a function of $P$, the typical pulse repetition period, under the assumption that any emitted pulse would exceed the minimum pulse sensitivity of the instrument. The results are replicated in Figure~\ref{fig:haystack-fig}. We emphasize, however, that the minimum pulse sensitivity of the VERITAS observations ($\gtrsim3\U{ph}\UU{m}{-2}$) is much better than that of the Harvard experiment ($\gtrsim100\U{ph}\UU{m}{-2}$).

We have applied a similar methodology to the sum of both VERITAS datasets described in this paper, with an observed sample of 247 unique stellar targets, and a total observation time of $140\U{hr}$. Figure~\ref{fig:haystack-fig} also demonstrates the potential survey sensitivity that can be achieved if we still obtain no candidate events after removing the constraint that pulses must be associated with a pre-defined location from the Breakthrough Listen target list.  This requires some additional analysis development, to further reduce the remaining background, but is realistically achievable in the near future. For this calculation, we assume a typical stellar density of $0.1\U{stars}\UU{pc}{-3}$ and a maximum range of $1\U{kpc}$, corresponding to $4\times10^8$ stars over the whole sky --- similar to the values used for calculating the Harvard All-Sky limits \citep{curtis_2013}. Using only the $140\U{hr}$ dataset considered in this paper, this search would lower the upper-limit on the fraction of stars with transmitters by roughly five orders of magnitude, corresponding to the ratio of the number of stars searched between targeted and non-targeted techniques. Applying the same approach to the entire VERITAS archive of 18,176 hours would further reduce the minimum upper limit, and extend the search sensitivity to much longer pulse transmission periods.  

\section{Conclusions and Prospects}

VERITAS is not alone in searching for optical technosignatures. Table 1 in \citet{2016ApJ...825L...5S} summarized the capabilities of optical technosignature searches at the time of the VERITAS analysis of KIC~462852. Since then, there have been numerous developments in the field. The Near-InfraRed Optical SETI instrumentation on a 1-meter telescope at the Lick Observatory has been used to conduct a survey of 1280 celestial objects in the near-infrared (950–$1650\U{nm}$), sensitive to pulses with durations of $<50\U{ns}$ \citep{2019AJ....158..203M}. \cite{2017AJ....153..251T} conducted a survey of 5600 FGKM stars using the Keck $10\U{m}$ telescope, searching for a spectral (not temporal) signature of laser emission. An all-sky instrument called PANOSETI is under development \citep{2020SPIE11447E..7GL, 2022SPIE12184E..8BM}, and will soon provide nightly all-sky coverage from two sites. Cherenkov telescopes also have an important role to play in future developments. The TAIGA-HiSCORE wide-aperture
Cherenkov array, consisting of 100, $0.5\UU{m}{2}$ telescopes spread over $1\UU{km}{2}$ with a field of view of $0.6\U{sr}$, has searched for nanosecond optical transients and detected pulsed emission from the CALIPSO satellite \citep{2021arXiv210909637P}. In the coming decade, the Cherenkov Telescope Array (CTA) will provide unprecedented telescope light collecting area, exceeding the mirror area of all of the world's large optical telescopes combined \citep{2019scta.book.....C}. It will have the capability to conduct nanosecond optical pulse searches similar to VERITAS, with much greater sensitivity and stricter background rejection. As we have shown here, verification and calibration of this capability with satellite-based lasers will be an important component of this program, as will considerations of the telescope optical performance and trigger system design.

\begin{acknowledgments}
This research is supported by grants from the U.S. Department of Energy Office of Science, the U.S. National Science Foundation and the Smithsonian Institution, by NSERC in Canada, and by the Helmholtz Association in Germany. This research used resources provided by the Open Science Grid, which is supported by the National Science Foundation and the U.S. Department of Energy's Office of Science, and resources of the National Energy Research Scientific Computing Center (NERSC), a U.S. Department of Energy Office of Science User Facility operated under Contract No. DE-AC02-05CH11231. We thank the Breakthrough Prize Foundation and the University of California, Berkeley, for their support. We also acknowledge the excellent work of the technical support staff at the Fred Lawrence Whipple Observatory and at the collaborating institutions in the construction and operation of the instrument.
\end{acknowledgments}

\textit{\large Services}: Celestrak, NASA Exoplanet Archive, TeVCat, the SIMBAD database, the ViZieR catalog  access tool

\software{ROOT \citep{2009CoPhC.180.2499A}, Astropy \citep{2022ApJ...935..167A}, Astroquery \citep{2019AJ....157...98G}, BeautifulSoup, GeoPy, Matplotlib \citep{Hunter:2007}, Mechanize, NetworkX \citep{SciPyProceedings_11}, Numpy \citep{harris2020array}, Pandas \citep{mckinney-proc-scipy-2010}, PyTeVCat, Seaborn \citep{Waskom2021}, Skyfield \citep{Rhodes_Skyfield_Generate_high_2020}}

\bibliography{bibli}

\begin{thebibliography}{}
\expandafter\ifx\csname natexlab\endcsname\relax\def\natexlab#1{#1}\fi
\providecommand{\url}[1]{\href{#1}{#1}}
\providecommand{\dodoi}[1]{doi:~\href{http://doi.org/#1}{\nolinkurl{#1}}}
\providecommand{\doeprint}[1]{\href{http://ascl.net/#1}{\nolinkurl{http://ascl.net/#1}}}
\providecommand{\doarXiv}[1]{\href{https://arxiv.org/abs/#1}{\nolinkurl{https://arxiv.org/abs/#1}}}

\bibitem[{{Abeysekara} {et~al.}(2016){Abeysekara}, {Archambault}, {Archer},
  {Benbow}, {Bird}, {Buchovecky}, {Buckley}, {Byrum}, {Cardenzana}, {Cerruti},
  {Chen}, {Christiansen}, {Ciupik}, {Cui}, {Dickinson}, {Eisch}, {Errando},
  {Falcone}, {Fegan}, {Feng}, {Finley}, {Fleischhack}, {Fortin}, {Fortson},
  {Furniss}, {Gillanders}, {Griffin}, {Grube}, {Gyuk}, {H{\"u}tten},
  {H{\r{a}}kansson}, {Hanna}, {Holder}, {Humensky}, {Johnson}, {Kaaret}, {Kar},
  {Kelley-Hoskins}, {Kertzman}, {Kieda}, {Krause}, {Krennrich}, {Kumar},
  {Lang}, {Lin}, {Maier}, {McArthur}, {McCann}, {Meagher}, {Moriarty},
  {Mukherjee}, {Nieto}, {O'Brien}, {O'Faol{\'a}in de Bhr{\'o}ithe}, {Ong},
  {Otte}, {Park}, {Perkins}, {Petrashyk}, {Pohl}, {Popkow}, {Pueschel},
  {Quinn}, {Ragan}, {Ratliff}, {Reynolds}, {Richards}, {Roache}, {Santander},
  {Sembroski}, {Shahinyan}, {Staszak}, {Telezhinsky}, {Tucci}, {Tyler},
  {Vincent}, {Wakely}, {Weiner}, {Weinstein}, {Williams}, \&
  {Zitzer}}]{2016ApJ...818L..33A}
{Abeysekara}, A.~U., {Archambault}, S., {Archer}, A., {et~al.} 2016, \apjl,
  818, L33, \dodoi{10.3847/2041-8205/818/2/L33}

\bibitem[{{Antcheva} {et~al.}(2009){Antcheva}, {Ballintijn}, {Bellenot},
  {Biskup}, {Brun}, {Buncic}, {Canal}, {Casadei}, {Couet}, {Fine}, {Franco},
  {Ganis}, {Gheata}, {Maline}, {Goto}, {Iwaszkiewicz}, {Kreshuk}, {Segura},
  {Maunder}, {Moneta}, {Naumann}, {Offermann}, {Onuchin}, {Panacek},
  {Rademakers}, {Russo}, \& {Tadel}}]{2009CoPhC.180.2499A}
{Antcheva}, I., {Ballintijn}, M., {Bellenot}, B., {et~al.} 2009, Computer
  Physics Communications, 180, 2499, \dodoi{10.1016/j.cpc.2009.08.005}

\bibitem[{{Armada} {et~al.}(2005){Armada}, {Cortina}, \&
  {Martinez}}]{2005neeu.conf..307A}
{Armada}, A., {Cortina}, J., \& {Martinez}, M. 2005, in Neutrinos and Explosive
  Events in the Universe, ed. M.~M. {Shapiro}, T.~{Stanev}, \& J.~P. {Wefel},
  Vol. 209, 307

\bibitem[{{Astropy Collaboration} {et~al.}(2022){Astropy Collaboration},
  {Price-Whelan}, {Lim}, {Earl}, {Starkman}, {Bradley}, {Shupe}, {Patil},
  {Corrales}, {Brasseur}, {N{\"o}the}, {Donath}, {Tollerud}, {Morris},
  {Ginsburg}, {Vaher}, {Weaver}, {Tocknell}, {Jamieson}, {van Kerkwijk},
  {Robitaille}, {Merry}, {Bachetti}, {G{\"u}nther}, {Aldcroft},
  {Alvarado-Montes}, {Archibald}, {B{\'o}di}, {Bapat}, {Barentsen},
  {Baz{\'a}n}, {Biswas}, {Boquien}, {Burke}, {Cara}, {Cara}, {Conroy},
  {Conseil}, {Craig}, {Cross}, {Cruz}, {D'Eugenio}, {Dencheva}, {Devillepoix},
  {Dietrich}, {Eigenbrot}, {Erben}, {Ferreira}, {Foreman-Mackey}, {Fox},
  {Freij}, {Garg}, {Geda}, {Glattly}, {Gondhalekar}, {Gordon}, {Grant},
  {Greenfield}, {Groener}, {Guest}, {Gurovich}, {Handberg}, {Hart},
  {Hatfield-Dodds}, {Homeier}, {Hosseinzadeh}, {Jenness}, {Jones}, {Joseph},
  {Kalmbach}, {Karamehmetoglu}, {Ka{\l}uszy{\'n}ski}, {Kelley}, {Kern},
  {Kerzendorf}, {Koch}, {Kulumani}, {Lee}, {Ly}, {Ma}, {MacBride}, {Maljaars},
  {Muna}, {Murphy}, {Norman}, {O'Steen}, {Oman}, {Pacifici}, {Pascual},
  {Pascual-Granado}, {Patil}, {Perren}, {Pickering}, {Rastogi}, {Roulston},
  {Ryan}, {Rykoff}, {Sabater}, {Sakurikar}, {Salgado}, {Sanghi}, {Saunders},
  {Savchenko}, {Schwardt}, {Seifert-Eckert}, {Shih}, {Jain}, {Shukla}, {Sick},
  {Simpson}, {Singanamalla}, {Singer}, {Singhal}, {Sinha}, {Sip{\H{o}}cz},
  {Spitler}, {Stansby}, {Streicher}, {{\v{S}}umak}, {Swinbank}, {Taranu},
  {Tewary}, {Tremblay}, {Val-Borro}, {Van Kooten}, {Vasovi{\'c}}, {Verma}, {de
  Miranda Cardoso}, {Williams}, {Wilson}, {Winkel}, {Wood-Vasey}, {Xue},
  {Yoachim}, {Zhang}, {Zonca}, \& {Astropy Project
  Contributors}}]{2022ApJ...935..167A}
{Astropy Collaboration}, {Price-Whelan}, A.~M., {Lim}, P.~L., {et~al.} 2022,
  \apj, 935, 167, \dodoi{10.3847/1538-4357/ac7c74}

\bibitem[{{Bracewell}(1960)}]{1960Natur.186..670B}
{Bracewell}, R.~N. 1960, Nature, 186, 670, \dodoi{10.1038/186670a0}

\bibitem[{{Canestrari} {et~al.}(2010){Canestrari}, {Motta}, {Pareschi},
  {Basso}, {Doro}, {Giro}, \& {Lessio}}]{2010SPIE.7739E..0HC}
{Canestrari}, R., {Motta}, G., {Pareschi}, G., {et~al.} 2010, in Society of
  Photo-Optical Instrumentation Engineers (SPIE) Conference Series, Vol. 7739,
  Modern Technologies in Space- and Ground-based Telescopes and
  Instrumentation, ed. E.~{Atad-Ettedgui} \& D.~{Lemke}, 77390H,
  \dodoi{10.1117/12.857268}

\bibitem[{{Cherenkov Telescope Array Consortium} {et~al.}(2019){Cherenkov
  Telescope Array Consortium}, {Acharya}, {Agudo}, {Al Samarai}, {Alfaro},
  {Alfaro}, {Alispach}, {Alves Batista}, {Amans}, {Amato}, {Ambrosi},
  {Antolini}, {Antonelli}, {Aramo}, {Araya}, {Armstrong}, {Arqueros},
  {Arrabito}, {Asano}, {Ashley}, {Backes}, {Balazs}, {Balbo}, {Ballester},
  {Ballet}, {Bamba}, {Barkov}, {Barres de Almeida}, {Barrio}, {Bastieri},
  {Becherini}, {Belfiore}, {Benbow}, {Berge}, {Bernardini}, {Bernardini},
  {Bernardos}, {Bernl{\"o}hr}, {Bertucci}, {Biasuzzi}, {Bigongiari}, {Biland},
  {Bissaldi}, {Biteau}, {Blanch}, {Blazek}, {Boisson}, {Bolmont}, {Bonanno},
  {Bonardi}, {Bonavolont{\`a}}, {Bonnoli}, {Bosnjak}, {B{\"o}ttcher},
  {Braiding}, {Bregeon}, {Brill}, {Brown}, {Brun}, {Brunetti}, {Buanes},
  {Buckley}, {Bugaev}, {B{\"u}hler}, {Bulgarelli}, {Bulik}, {Burton},
  {Burtovoi}, {Busetto}, {Canestrari}, {Capalbi}, {Capitanio}, {Caproni},
  {Caraveo}, {C{\'a}rdenas}, {Carlile}, {Carosi}, {Carqu{\'\i}n}, {Carr},
  {Casanova}, {Cascone}, {Catalani}, {Catalano}, {Cauz}, {Cerruti}, {Chadwick},
  {Chaty}, {Chaves}, {Chen}, {Chen}, {Chernyakova}, {Chikawa}, {Christov},
  {Chudoba}, {Cie{\'s}lar}, {Coco}, {Colafrancesco}, {Colin}, {Conforti},
  {Connaughton}, {Conrad}, {Contreras}, {Cortina}, {Costa}, {Costantini},
  {Cotter}, {Covino}, {Crocker}, {Cuadra}, {Cuevas}, {Cumani}, {D'A{\`\i}},
  {D'Ammando}, {D'Avanzo}, {D'Urso}, {Daniel}, {Davids}, {Dawson}, {Dazzi}, {De
  Angelis}, {de C{\'a}ssia dos Anjos}, {De Cesare}, {De Franco}, {de Gouveia
  Dal Pino}, {de la Calle}, {de los Reyes Lopez}, {De Lotto}, {De Luca}, {De
  Lucia}, {de Naurois}, {de O{\~n}a Wilhelmi}, {De Palma}, {De Persio}, {de
  Souza}, {Deil}, {Del Santo}, {Delgado}, {della Volpe}, {Di Girolamo}, {Di
  Pierro}, {Di Venere}, {D{\'\i}az}, {Dib}, {Diebold}, {Djannati-Ata{\"\i}},
  {Dom{\'\i}nguez}, {Dominis Prester}, {Dorner}, {Doro}, {Drass}, {Dravins},
  {Dubus}, {Dwarkadas}, {Ebr}, {Eckner}, {Egberts}, {Einecke}, {Ekoume},
  {Els{\"a}sser}, {Ernenwein}, {Espinoza}, {Evoli}, {Fairbairn},
  {Falceta-Goncalves}, {Falcone}, {Farnier}, {Fasola}, {Fedorova}, {Fegan},
  {Fernandez-Alonso}, {Fern{\'a}ndez-Barral}, {Ferrand}, {Fesquet},
  {Filipovic}, {Fioretti}, {Fontaine}, {Fornasa}, {Fortson}, {Freixas
  Coromina}, {Fruck}, {Fujita}, {Fukazawa}, {Funk}, {F{\"u}{\ss}ling},
  {Gabici}, {Gadola}, {Gallant}, {Garcia}, {Garcia L{\'o}pez}, {Garczarczyk},
  {Gaskins}, {Gasparetto}, {Gaug}, {Gerard}, {Giavitto}, {Giglietto}, {Giommi},
  {Giordano}, {Giro}, {Giroletti}, {Giuliani}, {Glicenstein}, {Gnatyk},
  {Godinovic}, {Goldoni}, {G{\'o}mez-Vargas}, {Gonz{\'a}lez}, {Gonz{\'a}lez},
  {G{\"o}tz}, {Graham}, {Grandi}, {Granot}, {Green}, {Greenshaw}, {Griffiths},
  {Gunji}, {Hadasch}, {Hara}, {Hardcastle}, {Hassan}, {Hayashi}, {Hayashida},
  {Heller}, {Helo}, {Hermann}, {Hinton}, {Hnatyk}, {Hofmann}, {Holder},
  {Horan}, {H{\"o}randel}, {Horns}, {Horvath}, {Hovatta}, {Hrabovsky},
  {Hrupec}, {Humensky}, {H{\"u}tten}, {Iarlori}, {Inada}, {Inome}, {Inoue},
  {Inoue}, {Inoue}, {Iocco}, {Ioka}, {Iori}, {Ishio}, {Iwamura}, {Jamrozy},
  {Janecek}, {Jankowsky}, {Jean}, {Jung-Richardt}, {Jurysek}, {Kaaret},
  {Karkar}, {Katagiri}, {Katz}, {Kawanaka}, {Kazanas}, {Kh{\'e}lifi}, {Kieda},
  {Kimeswenger}, {Kimura}, {Kisaka}, {Knapp}, {Kn{\"o}dlseder}, {Koch},
  {Kohri}, {Komin}, {Kosack}, {Kraus}, {Krause}, {Krau{\ss}}, {Kubo}, {Kukec
  Mezek}, {Kuroda}, {Kushida}, {La Palombara}, {Lamanna}, {Lang}, {Lapington},
  {Le Blanc}, {Leach}, {Lees}, {Lefaucheur}, {Leigui de Oliveira}, {Lenain},
  {Lico}, {Limon}, {Lindfors}, {Lohse}, {Lombardi}, {Longo}, {L{\'o}pez},
  {L{\'o}pez-Coto}, {Lu}, {Lucarelli}, {Luque-Escamilla}, {Lyard}, {Maccarone},
  {Maier}, {Majumdar}, {Malaguti}, {Mandat}, {Maneva}, {Manganaro}, {Mangano},
  {Marcowith}, {Mar{\'\i}n}, {Markoff}, {Mart{\'\i}}, {Martin},
  {Mart{\'\i}nez}, {Mart{\'\i}nez}, {Masetti}, {Masuda}, {Maurin}, {Maxted},
  {Mazin}, {Medina}, {Melandri}, {Mereghetti}, {Meyer}, {Minaya}, {Mirabal},
  {Mirzoyan}, {Mitchell}, {Mizuno}, {Moderski}, {Mohammed}, {Mohrmann},
  {Montaruli}, {Moralejo}, {Morcuende-Parrilla}, {Mori}, {Morlino}, {Morris},
  {Morselli}, {Moulin}, {Mukherjee}, {Mundell}, {Murach}, {Muraishi}, {Murase},
  {Nagai}, {Nagataki}, {Nagayoshi}, {Naito}, {Nakamori}, {Nakamura}, {Niemiec},
  {Nieto}, {Niko{\l}ajuk}, {Nishijima}, {Noda}, {Nosek}, {Novosyadlyj},
  {Nozaki}, {O'Brien}, {Oakes}, {Ohira}, {Ohishi}, {Ohm}, {Okazaki}, {Okumura},
  {Ong}, {Orienti}, {Orito}, {Osborne}, {Ostrowski}, {Otte}, {Oya}, {Padovani},
  {Paizis}, {Palatiello}, {Palatka}, {Paoletti}, {Paredes}, {Pareschi},
  {Parsons}, {Pe'er}, {Pech}, {Pedaletti}, {Perri}, {Persic}, {Petrashyk},
  {Petrucci}, {Petruk}, {Peyaud}, {Pfeifer}, {Piano}, {Pisarski}, {Pita},
  {Pohl}, {Polo}, {Pozo}, {Prandini}, {Prast}, {Principe}, {Prokhorov},
  {Prokoph}, {Prouza}, {P{\"u}hlhofer}, {Punch}, {P{\"u}rckhauer}, {Queiroz},
  {Quirrenbach}, {Rain{\`o}}, {Razzaque}, {Reimer}, {Reimer}, {Reisenegger},
  {Renaud}, {Rezaeian}, {Rhode}, {Ribeiro}, {Rib{\'o}}, {Richtler}, {Rico},
  {Rieger}, {Riquelme}, {Rivoire}, {Rizi}, {Rodriguez}, {Rodriguez Fernandez},
  {Rodr{\'\i}guez V{\'a}zquez}, {Rojas}, {Romano}, {Romeo}, {Rosado}, {Rovero},
  {Rowell}, {Rudak}, {Rugliancich}, {Rulten}, {Sadeh}, {Safi-Harb}, {Saito},
  {Sakaki}, {Sakurai}, {Salina}, {S{\'a}nchez-Conde}, {Sandaker}, {Sandoval},
  {Sangiorgi}, {Sanguillon}, {Sano}, {Santander}, {Sarkar}, {Satalecka},
  {Saturni}, {Schioppa}, {Schlenstedt}, {Schneider}, {Schoorlemmer},
  {Schovanek}, {Schulz}, {Schussler}, {Schwanke}, {Sciacca}, {Scuderi},
  {Seitenzahl}, {Semikoz}, {Sergijenko}, {Servillat}, {Shalchi}, {Shellard},
  {Sidoli}, {Siejkowski}, {Sillanp{\"a}{\"a}}, {Sironi}, {Sitarek}, {Sliusar},
  {Slowikowska}, {Sol}, {Stamerra}, {Stani{\v{c}}}, {Starling}, {Stawarz},
  {Stefanik}, {Stephan}, {Stolarczyk}, {Stratta}, {Straumann}, {Suomijarvi},
  {Supanitsky}, {Tagliaferri}, {Tajima}, {Tavani}, {Tavecchio}, {Tavernet},
  {Tayabaly}, {Tejedor}, {Temnikov}, {Terada}, {Terrier}, {Terzic}, {Teshima},
  {Testa}, {Thoudam}, {Tian}, {Tibaldo}, {Tluczykont}, {Todero Peixoto},
  {Tokanai}, {Tomastik}, {Tonev}, {Tornikoski}, {Torres}, {Torresi}, {Tosti},
  {Tothill}, {Tovmassian}, {Travnicek}, {Trichard}, {Trifoglio}, {Troyano
  Pujadas}, {Tsujimoto}, {Umana}, {Vagelli}, {Vagnetti}, {Valentino},
  {Vallania}, {Valore}, {van Eldik}, {Vandenbroucke}, {Varner}, {Vasileiadis},
  {Vassiliev}, {V{\'a}zquez Acosta}, {Vecchi}, {Vega}, {Vercellone}, {Veres},
  {Vergani}, {Verzi}, {Vettolani}, {Viana}, {Vigorito}, {Villanueva}, {Voelk},
  {Vollhardt}, {Vorobiov}, {Vrastil}, {Vuillaume}, {Wagner}, {Wagner},
  {Walter}, {Ward}, {Warren}, {Watson}, {Werner}, {White}, {White},
  {Wierzcholska}, {Wilcox}, {Will}, {Williams}, {Wischnewski}, {Wood},
  {Yamamoto}, {Yamazaki}, {Yanagita}, {Yang}, {Yoshida}, {Yoshiike},
  {Yoshikoshi}, {Zacharias}, {Zaharijas}, {Zampieri}, {Zandanel}, {Zanin},
  {Zavrtanik}, {Zavrtanik}, {Zdziarski}, {Zech}, {Zechlin}, {Zhdanov},
  {Ziegler}, \& {Zorn}}]{2019scta.book.....C}
{Cherenkov Telescope Array Consortium}, {Acharya}, B.~S., {Agudo}, I., {et~al.}
  2019, {Science with the Cherenkov Telescope Array}, \dodoi{10.1142/10986}

\bibitem[{{Cocconi} \& {Morrison}(1959)}]{1959Natur.184..844C}
{Cocconi}, G., \& {Morrison}, P. 1959, \nat, 184, 844, \dodoi{10.1038/184844a0}

\bibitem[{{Cogan}(2008)}]{2008ICRC....3.1385C}
{Cogan}, P. 2008, in International Cosmic Ray Conference, Vol.~3, International
  Cosmic Ray Conference, 1385--1388.
\newblock \doarXiv{0709.4233}

\bibitem[{{Covault}(2001)}]{2001SPIE.4273..161C}
{Covault}, C.~E. 2001, in Society of Photo-Optical Instrumentation Engineers
  (SPIE) Conference Series, Vol. 4273, The Search for Extraterrestrial
  Intelligence (SETI) in the Optical Spectrum III, ed. S.~A. {Kingsley} \&
  R.~{Bhathal}, 161--172, \dodoi{10.1117/12.435374}

\bibitem[{{Davies} \& {Cotton}(1957)}]{1957SoEn....1...16D}
{Davies}, J.~M., \& {Cotton}, E.~S. 1957, Solar Energy, 1, 16,
  \dodoi{10.1016/0038-092X(57)90116-0}

\bibitem[{{Dyson}(1960)}]{1960Sci...131.1667D}
{Dyson}, F.~J. 1960, Science, 131, 1667, \dodoi{10.1126/science.131.3414.1667}

\bibitem[{{Eichler} \& {Beskin}(2001)}]{2001AsBio...1..489E}
{Eichler}, D., \& {Beskin}, G. 2001, Astrobiology, 1, 489,
  \dodoi{10.1089/153110701753593892}

\bibitem[{{Franz} {et~al.}(2022){Franz}, {Croft}, {Siemion}, {Traas},
  {Brzycki}, {Gajjar}, {Isaacson}, {Lebofsky}, {MacMahon}, {Price}, {Sheikh},
  {DeMarines}, {Drew}, \& {Worden}}]{Franz:2022}
{Franz}, N., {Croft}, S., {Siemion}, A. P.~V., {et~al.} 2022, \aj, 163, 104,
  \dodoi{10.3847/1538-3881/ac46c9}

\bibitem[{{Gajjar} {et~al.}(2019){Gajjar}, {Siemion}, {Croft}, {Brzycki},
  {Burgay}, {Carozzi}, {Concu}, {Czech}, {DeBoer}, {DeMarines}, {Drew},
  {Enriquez}, {Fawcett}, {Gallagher}, {Gerret}, {Gizani}, {Hellbourg},
  {Holder}, {Isaacson}, {Kudale}, {Lacki}, {Lebofsky}, {Li}, {MacMahon},
  {McCauley}, {Melis}, {Molinari}, {Murphy}, {Perrodin}, {Pilia}, {Price},
  {Webb}, {Werthimer}, {Williams}, {Worden}, {Zarka}, \&
  {Zhang}}]{2019BAAS...51g.223G}
{Gajjar}, V., {Siemion}, A., {Croft}, S., {et~al.} 2019, in Bulletin of the
  American Astronomical Society, Vol.~51, 223.
\newblock \doarXiv{1907.05519}

\bibitem[{{Gingrich} {et~al.}(2005){Gingrich}, {Boone}, {Bramel}, {Carson},
  {Covault}, {Fortin}, {Hanna}, {Hinton}, {Jarvis}, {Kildea}, {Lindner},
  {Mueller}, {Mukherjee}, {Ong}, {Ragan}, {Scalzo}, {Theoret}, {Williams}, \&
  {Zweerink}}]{2005ITNS...52.2977G}
{Gingrich}, D.~M., {Boone}, L.~M., {Bramel}, D., {et~al.} 2005, IEEE
  Transactions on Nuclear Science, 52, 2977, \dodoi{10.1109/TNS.2005.855705}

\bibitem[{{Ginsburg} {et~al.}(2019){Ginsburg}, {Sip{\H o}cz}, {Brasseur},
  {Cowperthwaite}, {Craig}, {Deil}, {Guillochon}, {Guzman}, {Liedtke}, {Lian
  Lim}, {Lockhart}, {Mommert}, {Morris}, {Norman}, {Parikh}, {Persson},
  {Robitaille}, {Segovia}, {Singer}, {Tollerud}, {de Val-Borro}, {Valtchanov},
  {Woillez}, {The Astroquery collaboration}, \& {a subset of the astropy
  collaboration}}]{2019AJ....157...98G}
{Ginsburg}, A., {Sip{\H o}cz}, B.~M., {Brasseur}, C.~E., {et~al.} 2019, \aj,
  157, 98, \dodoi{10.3847/1538-3881/aafc33}

\bibitem[{Hagberg {et~al.}(2008)Hagberg, Schult, \&
  Swart}]{SciPyProceedings_11}
Hagberg, A.~A., Schult, D.~A., \& Swart, P.~J. 2008, in Proceedings of the 7th
  Python in Science Conference, ed. G.~Varoquaux, T.~Vaught, \& J.~Millman,
  Pasadena, CA USA, 11 -- 15

\bibitem[{{Hanna}(2008)}]{2008ICRC....3.1417H}
{Hanna}, D. 2008, in International Cosmic Ray Conference, Vol.~3, International
  Cosmic Ray Conference, 1417--1420.
\newblock \doarXiv{0709.4479}

\bibitem[{{Hanna} {et~al.}(2010){Hanna}, {McCann}, {McCutcheon}, \&
  {Nikkinen}}]{2010NIMPA.612..278H}
{Hanna}, D., {McCann}, A., {McCutcheon}, M., \& {Nikkinen}, L. 2010, NIMPA,
  612, 278, \dodoi{10.1016/j.nima.2009.10.107}

\bibitem[{{Hanna} {et~al.}(2009)}]{2009AsBio...9..345H}
{Hanna}, D.~S., {et~al.} 2009, Astrobiology, 9, 345,
  \dodoi{10.1089/ast.2008.0256}

\bibitem[{Harris {et~al.}(2020)Harris, Millman, van~der Walt, Gommers,
  Virtanen, Cournapeau, Wieser, Taylor, Berg, Smith, Kern, Picus, Hoyer, van
  Kerkwijk, Brett, Haldane, del R{\'{i}}o, Wiebe, Peterson,
  G{\'{e}}rard-Marchant, Sheppard, Reddy, Weckesser, Abbasi, Gohlke, \&
  Oliphant}]{harris2020array}
Harris, C.~R., Millman, K.~J., van~der Walt, S.~J., {et~al.} 2020, Nature, 585,
  357, \dodoi{10.1038/s41586-020-2649-2}

\bibitem[{{Heller} \& {Pudritz}(2016)}]{2016AsBio..16..259H}
{Heller}, R., \& {Pudritz}, R.~E. 2016, AsBio, 16, 259,
  \dodoi{10.1089/ast.2015.1358}

\bibitem[{{Hillas}(1985)}]{1985ICRC....3..445H}
{Hillas}, A.~M. 1985, in International Cosmic Ray Conference, Vol.~3, 19th
  International Cosmic Ray Conference (ICRC19), Volume 3, 445

\bibitem[{{Hippke}(2018)}]{2018JApA...39...73H}
{Hippke}, M. 2018, JApA, 39, 73, \dodoi{10.1007/s12036-018-9566-x}

\bibitem[{Holder(2021)}]{JHolder2021Chapter6WSPCHandbook}
Holder, J. 2021, Atmospheric Cherenkov Gamma-Ray Telescopes, 2nd edn., World
  Scientific Series in Astrophysics (World Scientific), 117--136,
  \dodoi{10.1142/9789811203817_0006}

\bibitem[{{Holder} {et~al.}(2005){Holder}, {Ashworth}, {LeBohec}, {Rose}, \&
  {Weekes}}]{2005ICRC....5..387H}
{Holder}, J., {Ashworth}, P., {LeBohec}, S., {Rose}, H.~J., \& {Weekes}, T.~C.
  2005, in International Cosmic Ray Conference, Vol.~5, 29th International
  Cosmic Ray Conference (ICRC29), Volume 5, 387.
\newblock \doarXiv{astro-ph/0506758}

\bibitem[{{Holder} {et~al.}(2006){Holder}, {Atkins}, {Badran}, {Blaylock},
  {Bradbury}, {Buckley}, {Byrum}, {Carter-Lewis}, {Celik}, {Chow}, {Cogan},
  {Cui}, {Daniel}, {de la Calle Perez}, {Dowdall}, {Dowkontt}, {Duke},
  {Falcone}, {Fegan}, {Finley}, {Fortin}, {Fortson}, {Gibbs}, {Gillanders},
  {Glidewell}, {Grube}, {Gutierrez}, {Gyuk}, {Hall}, {Hanna}, {Hays}, {Horan},
  {Hughes}, {Humensky}, {Imran}, {Jung}, {Kaaret}, {Kenny}, {Kieda}, {Kildea},
  {Knapp}, {Krawczynski}, {Krennrich}, {Lang}, {LeBohec}, {Linton}, {Little},
  {Maier}, {Manseri}, {Milovanovic}, {Moriarty}, {Mukherjee}, {Ogden}, {Ong},
  {Petry}, {Perkins}, {Pizlo}, {Pohl}, {Quinn}, {Ragan}, {Reynolds}, {Roache},
  {Rose}, {Schroedter}, {Sembroski}, {Sleege}, {Steele}, {Swordy}, {Syson},
  {Toner}, {Valcarcel}, {Vassiliev}, {Wakely}, {Weekes}, {White}, {Williams},
  \& {Wagner}}]{2006APh....25..391H}
{Holder}, J., {Atkins}, R.~W., {Badran}, H.~M., {et~al.} 2006, Astroparticle
  Physics, 25, 391, \dodoi{10.1016/j.astropartphys.2006.04.002}

\bibitem[{{Horowitz} {et~al.}(2001){Horowitz}, {Coldwell}, {Howard}, {Latham},
  {Stefanik}, {Wolff}, \& {Zajac}}]{2001SPIE.4273..119H}
{Horowitz}, P., {Coldwell}, C.~M., {Howard}, A.~B., {et~al.} 2001, in Society
  of Photo-Optical Instrumentation Engineers (SPIE) Conference Series, Vol.
  4273, The Search for Extraterrestrial Intelligence (SETI) in the Optical
  Spectrum III, ed. S.~A. {Kingsley} \& R.~{Bhathal}, 119--127,
  \dodoi{10.1117/12.435364}

\bibitem[{{Howard} {et~al.}(2007){Howard}, {Horowitz}, {Mead}, {Sreetharan},
  {Gallicchio}, {Howard}, {Coldwell}, {Zajac}, \&
  {Sliski}}]{2007AcAau..61...78H}
{Howard}, A., {Horowitz}, P., {Mead}, C., {et~al.} 2007, Acta Astronautica, 61,
  78, \dodoi{10.1016/j.actaastro.2007.01.038}

\bibitem[{{Howard}(2006)}]{2006PhDT........58H}
{Howard}, A.~W. 2006, PhD thesis, Harvard University

\bibitem[{{Howard} {et~al.}(2004){Howard}, {Horowitz}, {Wilkinson}, {Coldwell},
  {Groth}, {Jarosik}, {Latham}, {Stefanik}, {Willman}, {Wolff}, \&
  {Zajac}}]{2004ApJ...613.1270H}
{Howard}, A.~W., {Horowitz}, P., {Wilkinson}, D.~T., {et~al.} 2004, \apj, 613,
  1270, \dodoi{10.1086/423300}

\bibitem[{Hunter(2007)}]{Hunter:2007}
Hunter, J.~D. 2007, Computing in Science \& Engineering, 9, 90,
  \dodoi{10.1109/MCSE.2007.55}

\bibitem[{{Isaacson} {et~al.}(2019){Isaacson}, {Siemion}, {Marcy}, {Hickish},
  {Price}, {Enriquez}, \& {Gizani}}]{Isaacson:2019}
{Isaacson}, H., {Siemion}, A. P.~V., {Marcy}, G.~W., {et~al.} 2019, \pasp, 131,
  014201, \dodoi{10.1088/1538-3873/aaeae0}

\bibitem[{{Isaacson} {et~al.}(2017{\natexlab{a}}){Isaacson}, {Siemion},
  {Marcy}, {Lebofsky}, {Price}, {MacMahon}, {Croft}, {DeBoer}, {Hickish},
  {Werthimer}, {Sheikh}, {Hellbourg}, \& {Enriquez}}]{Isaacson:2017}
---. 2017{\natexlab{a}}, \pasp, 129, 054501, \dodoi{10.1088/1538-3873/aa5800}

\bibitem[{{Isaacson} {et~al.}(2017{\natexlab{b}}){Isaacson}, {Siemion},
  {Marcy}, {Lebofsky}, {Price}, {MacMahon}, {Croft}, {DeBoer}, {Hickish},
  {Werthimer}, {Sheikh}, {Hellbourg}, \& {Enriquez}}]{2017PASP..129e4501I}
---. 2017{\natexlab{b}}, \pasp, 129, 054501, \dodoi{10.1088/1538-3873/aa5800}

\bibitem[{{Kieda}(2013)}]{2013ICRC...33.1124K}
{Kieda}, D.~B. 2013, in International Cosmic Ray Conference, Vol.~33,
  International Cosmic Ray Conference, 1124.
\newblock \doarXiv{1308.4849}

\bibitem[{{Krause} {et~al.}(2017){Krause}, {Pueschel}, \&
  {Maier}}]{2017APh....89....1K}
{Krause}, M., {Pueschel}, E., \& {Maier}, G. 2017, Astroparticle Physics, 89,
  1, \dodoi{10.1016/j.astropartphys.2017.01.004}

\bibitem[{{Lipman} {et~al.}(2019){Lipman}, {Isaacson}, {Siemion}, {Lebofsky},
  {Price}, {MacMahon}, {Croft}, {DeBoer}, {Hickish}, {Werthimer}, {Hellbourg},
  {Enriquez}, \& {Gizani}}]{Lipman:2019}
{Lipman}, D., {Isaacson}, H., {Siemion}, A. P.~V., {et~al.} 2019, \pasp, 131,
  034202, \dodoi{10.1088/1538-3873/aafe86}

\bibitem[{{Liu} {et~al.}(2020){Liu}, {Werthimer}, {Lee}, {Antonio}, {Aronson},
  {Brown}, {Drake}, {Howard}, {Horowitz}, {Maire}, {Raffanti}, {Stone},
  {Treffers}, \& {Wright}}]{2020SPIE11447E..7GL}
{Liu}, W., {Werthimer}, D., {Lee}, R., {et~al.} 2020, in Society of
  Photo-Optical Instrumentation Engineers (SPIE) Conference Series, Vol. 11447,
  Society of Photo-Optical Instrumentation Engineers (SPIE) Conference Series,
  114477G, \dodoi{10.1117/12.2561203}

\bibitem[{{Maier} \& {Holder}(2017)}]{2017ICRC...35..747M}
{Maier}, G., \& {Holder}, J. 2017, in International Cosmic Ray Conference, Vol.
  301, 35th International Cosmic Ray Conference (ICRC2017), 747.
\newblock \doarXiv{1708.04048}

\bibitem[{{Maire} {et~al.}(2019){Maire}, {Wright}, {Barrett}, {Dexter},
  {Dorval}, {Duenas}, {Drake}, {Hultgren}, {Isaacson}, {Marcy}, {Meyer},
  {Ramos}, {Shirman}, {Siemion}, {Stone}, {Tallis}, {Tellis}, {Treffers}, \&
  {Werthimer}}]{2019AJ....158..203M}
{Maire}, J., {Wright}, S.~A., {Barrett}, C.~T., {et~al.} 2019, \aj, 158, 203,
  \dodoi{10.3847/1538-3881/ab44d3}

\bibitem[{{Maire} {et~al.}(2022){Maire}, {Wright}, {Holder}, {Anderson},
  {Benbow}, {Brown}, {Cosens}, {Foote}, {Hanlon}, {Hervet}, {Horowitz},
  {Howard}, {Lee}, {Liu}, {Raffanti}, {Rault-Wang}, {Stone}, {Werthimer},
  {Wiley}, \& {Williams}}]{2022SPIE12184E..8BM}
{Maire}, J., {Wright}, S.~A., {Holder}, J., {et~al.} 2022, in Society of
  Photo-Optical Instrumentation Engineers (SPIE) Conference Series, Vol. 12184,
  Ground-based and Airborne Instrumentation for Astronomy IX, ed. C.~J.
  {Evans}, J.~J. {Bryant}, \& K.~{Motohara}, 121848B,
  \dodoi{10.1117/12.2630772}

\bibitem[{{McCann} {et~al.}(2010){McCann}, {Hanna}, {Kildea}, \&
  {McCutcheon}}]{2010APh....32..325M}
{McCann}, A., {Hanna}, D., {Kildea}, J., \& {McCutcheon}, M. 2010,
  Astroparticle Physics, 32, 325, \dodoi{10.1016/j.astropartphys.2009.10.001}

\bibitem[{Mead(2013)}]{curtis_2013}
Mead, C.~C. 2013, Doctoral dissertation, Harvard University.
\newblock \url{http://nrs.harvard.edu/urn-3:HUL.InstRepos:11158246}

\bibitem[{{Nagai} {et~al.}(2008){Nagai}, {McKay}, {Sleege}, \&
  {Petry}}]{2008ICRC....3.1437N}
{Nagai}, T., {McKay}, R., {Sleege}, G., \& {Petry}, D. 2008, in International
  Cosmic Ray Conference, Vol.~3, International Cosmic Ray Conference,
  1437--1440

\bibitem[{{Panov} {et~al.}(2021){Panov}, {Astapov}, {Awad}, {Beskin},
  {Bezyazeekov}, {Blank}, {Bonvech}, {Borodin}, {Bruckner}, {Budnev}, {Bulan},
  {Chernov}, {Chiavassa}, {Dyachok}, {Gafarov}, {Garmash}, {Grebenyuk},
  {Gress}, {Gress}, {Grinyuk}, {Grishin}, {Horns}, {Ivanova}, {Kalmykov},
  {Kindin}, {Kiryuhin}, {Kokoulin}, {Kompaniets}, {Korosteleva}, {Kozhin},
  {Kravchenko}, {Krivopalova}, {Kuzmichev}, {Kryukov}, {Lagutin}, {Lavrova},
  {Lemeshev}, {Lubsandorzhiev}, {Lubsandorzhiev}, {Lukanov}, {Mirgazov},
  {Mirzoyan}, {Monkhoev}, {Osipova}, {Pakhorukov}, {Pan}, {Pankov},
  {Petrukhin}, {Podgrudkov}, {Poleschuk}, {Popova}, {Porelli}, {Postnikov},
  {Prosin}, {Ptuskin}, {Pushnin}, {Raikin}, {Razumov}, {Rjabov}, {Rubtsov},
  {Sagan}, {Samoliga}, {Sidorenkov}, {Silaev}, {Silaev}, {Skurikhin},
  {Satyshev}, {Sokolov}, {Suvorkin}, {Sveshnikova}, {Tabolenko}, {Tanaev},
  {Tarashansky}, {Ternovoy}, {Tkachev}, {Tluczykont}, {Ushakov},
  {Vaidyanathan}, {Volchugov}, {Volkov}, {Voronin}, {Wischnewski}, {Yashin},
  {Zagorodnikov}, \& {Zhurov}}]{2021arXiv210909637P}
{Panov}, A.~D., {Astapov}, I.~I., {Awad}, A.~K., {et~al.} 2021, arXiv e-prints,
  arXiv:2109.09637.
\newblock \doarXiv{2109.09637}

\bibitem[{{Porelli} \& {Taiga Collaboration}(2022)}]{2022icrc.confE.876P}
{Porelli}, A., \& {Taiga Collaboration}. 2022, in 37th International Cosmic Ray
  Conference, 876, \dodoi{10.22323/1.395.0876}

\bibitem[{Rhodes(2020)}]{Rhodes_Skyfield_Generate_high_2020}
Rhodes, B. 2020, {Skyfield: Generate high precision research-grade positions
  for stars, planets, moons, and Earth satellites}, 1.17

\bibitem[{{Roache} {et~al.}(2008){Roache}, {Irvin}, {Perkins}, {Harris},
  {Falcone}, {Finley}, \& {Weeks}}]{2008ICRC....3.1397R}
{Roache}, E., {Irvin}, R., {Perkins}, J.~S., {et~al.} 2008, in International
  Cosmic Ray Conference, Vol.~3, International Cosmic Ray Conference,
  1397--1400

\bibitem[{{Schneider} {et~al.}(2010){Schneider}, {L{\'e}ger}, {Fridlund},
  {White}, {Eiroa}, {Henning}, {Herbst}, {Lammer}, {Liseau}, {Paresce},
  {Penny}, {Quirrenbach}, {R{\"o}ttgering}, {Selsis}, {Beichman}, {Danchi},
  {Kaltenegger}, {Lunine}, {Stam}, \& {Tinetti}}]{2010AsBio..10..121S}
{Schneider}, J., {L{\'e}ger}, A., {Fridlund}, M., {et~al.} 2010, Astrobiology,
  10, 121, \dodoi{10.1089/ast.2009.0371}

\bibitem[{{Schuetz} {et~al.}(2016){Schuetz}, {Vakoch}, {Shostak}, \&
  {Richards}}]{2016ApJ...825L...5S}
{Schuetz}, M., {Vakoch}, D.~A., {Shostak}, S., \& {Richards}, J. 2016, \apjl,
  825, L5, \dodoi{10.3847/2041-8205/825/1/L5}

\bibitem[{{Schwartz} \& {Townes}(1961)}]{1961Natur.190..205S}
{Schwartz}, R.~N., \& {Townes}, C.~H. 1961, \nat, 190, 205,
  \dodoi{10.1038/190205a0}

\bibitem[{{Sheikh} {et~al.}(2020){Sheikh}, {Siemion}, {Enriquez}, {Price},
  {Isaacson}, {Lebofsky}, {Gajjar}, \& {Kalas}}]{2020AJ....160...29S}
{Sheikh}, S.~Z., {Siemion}, A., {Enriquez}, J.~E., {et~al.} 2020, AJ, 160, 29,
  \dodoi{10.3847/1538-3881/ab9361}

\bibitem[{{Shepherd} {et~al.}(2005){Shepherd}, {Buckley}, {Celik}, {Holder},
  {LeBohec}, {Manseri}, {Pizlo}, \& {Roberts}}]{2005ICRC....5..427S}
{Shepherd}, N., {Buckley}, J.~H., {Celik}, O., {et~al.} 2005, in International
  Cosmic Ray Conference, Vol.~5, 29th International Cosmic Ray Conference
  (ICRC29), Volume 5, 427.
\newblock \doarXiv{astro-ph/0507083}

\bibitem[{{Siemion} {et~al.}(2015){Siemion}, {Benford}, {Cheng-Jin},
  {Chennamangalam}, {Cordes}, {Falcke}, {Garrington}, {Garrett}, {Gurvits},
  {Hoare}, {Korpela}, {Lazio}, {Messerschmitt}, {Morrison}, {O'Brien},
  {Paragi}, {Penny}, {Spitler}, {Tarter}, \& {Werthimer}}]{Siemion:2015}
{Siemion}, A., {Benford}, J., {Cheng-Jin}, J., {et~al.} 2015, in Advancing
  Astrophysics with the Square Kilometre Array (AASKA14), 116,
  \dodoi{10.22323/1.215.0116}

\bibitem[{{Sullivan} {et~al.}(1978){Sullivan}, {Brown}, \&
  {Wetherill}}]{1978Sci...199..377S}
{Sullivan}, W.~T., I., {Brown}, S., \& {Wetherill}, C. 1978, Science, 199, 377,
  \dodoi{10.1126/science.199.4327.377}

\bibitem[{{Tarter}(2003)}]{2003ESASP.539...31T}
{Tarter}, J. 2003, in ESA Special Publication, Vol. 539, Earths: DARWIN/TPF and
  the Search for Extrasolar Terrestrial Planets, ed. M.~{Fridlund},
  T.~{Henning}, \& H.~{Lacoste}, 31--38

\bibitem[{{Tellis} \& {Marcy}(2017)}]{2017AJ....153..251T}
{Tellis}, N.~K., \& {Marcy}, G.~W. 2017, \aj, 153, 251,
  \dodoi{10.3847/1538-3881/aa6d12}

\bibitem[{{Traas} {et~al.}(2021){Traas}, {Croft}, {Gajjar}, {Isaacson},
  {Lebofsky}, {MacMahon}, {Perez}, {Price}, {Sheikh}, {Siemion}, {Smith},
  {Drew}, \& {Worden}}]{Traas:2021}
{Traas}, R., {Croft}, S., {Gajjar}, V., {et~al.} 2021, \aj, 161, 286,
  \dodoi{10.3847/1538-3881/abf649}

\bibitem[{{Turnbull} \& {Tarter}(2003)}]{2003ApJS..145..181T}
{Turnbull}, M.~C., \& {Tarter}, J.~C. 2003, \apjs, 145, 181,
  \dodoi{10.1086/345779}

\bibitem[{Waskom(2021)}]{Waskom2021}
Waskom, M.~L. 2021, Journal of Open Source Software, 6, 3021,
  \dodoi{10.21105/joss.03021}

\bibitem[{{W}es {M}c{K}inney(2010)}]{mckinney-proc-scipy-2010}
{W}es {M}c{K}inney. 2010, in {P}roceedings of the 9th {P}ython in {S}cience
  {C}onference, ed. {S}t\'efan van~der {W}alt \& {J}arrod {M}illman, 56 -- 61,
  \dodoi{10.25080/Majora-92bf1922-00a}

\bibitem[{{Winker} {et~al.}(2009){Winker}, {Vaughan}, {Omar}, {Hu}, {Powell},
  {Liu}, {Hunt}, \& {Young}}]{2009JAtOT..26.2310W}
{Winker}, D.~M., {Vaughan}, M.~A., {Omar}, A., {et~al.} 2009, JAtOT, 26, 2310,
  \dodoi{10.1175/2009JTECHA1281.1}

\bibitem[{{Worden} {et~al.}(2017){Worden}, {Drew}, {Siemion}, {Werthimer},
  {DeBoer}, {Croft}, {MacMahon}, {Lebofsky}, {Isaacson}, {Hickish}, {Price},
  {Gajjar}, \& {Wright}}]{Worden:2017}
{Worden}, S.~P., {Drew}, J., {Siemion}, A., {et~al.} 2017, Acta Astronautica,
  139, 98, \dodoi{10.1016/j.actaastro.2017.06.0 08}

\bibitem[{Wright(2018)}]{Wright2018HandbookOfExoplanets}
Wright, J.~T. 2018, Exoplanets and SETI, ed. H.~J. Deeg \& J.~A. Belmonte
  (Cham: Springer International Publishing), 3405--3412,
  \dodoi{10.1007/978-3-319-55333-7_186}

\bibitem[{{Wright} {et~al.}(2018{\natexlab{a}}){Wright}, {Kanodia}, \&
  {Lubar}}]{2018AJ....156..260W}
{Wright}, J.~T., {Kanodia}, S., \& {Lubar}, E. 2018{\natexlab{a}}, \aj, 156,
  260, \dodoi{10.3847/1538-3881/aae099}

\bibitem[{{Wright} {et~al.}(2018{\natexlab{b}}){Wright}, {Sheikh}, {Alm{\'a}r},
  {Denning}, {Dick}, \& {Tarter}}]{2018arXiv180906857W}
{Wright}, J.~T., {Sheikh}, S., {Alm{\'a}r}, I., {et~al.} 2018{\natexlab{b}},
  arXiv e-prints, arXiv:1809.06857.
\newblock \doarXiv{1809.06857}

\bibitem[{{Wright} {et~al.}(2018{\natexlab{c}}){Wright}, {Horowitz}, {Maire},
  {Werthimer}, {Antonio}, {Aronson}, {Chaim-Weismann}, {Cosens}, {Drake},
  {Howard}, {Marcy}, {Raffanti}, {Siemion}, {Stone}, {Treffers}, \&
  {Uttamchandani}}]{2018SPIE10702E..5IW}
{Wright}, S.~A., {Horowitz}, P., {Maire}, J., {et~al.} 2018{\natexlab{c}}, in
  Society of Photo-Optical Instrumentation Engineers (SPIE) Conference Series,
  Vol. 10702, Ground-based and Airborne Instrumentation for Astronomy VII, ed.
  C.~J. {Evans}, L.~{Simard}, \& H.~{Takami}, 107025I,
  \dodoi{10.1117/12.2314268}

\bibitem[{{Zuckerman} {et~al.}(2023){Zuckerman}, {Ko}, {Isaacson}, {Croft},
  {Price}, {Lebofsky}, \& {Siemion}}]{2023arXiv230106971Z}
{Zuckerman}, A., {Ko}, Z., {Isaacson}, H., {et~al.} 2023, arXiv,
  arXiv:2301.06971.
\newblock \doarXiv{2301.06971}

\end{thebibliography}
\end{document}